\documentclass{pasj01}
\draft
\let\leftroot\relax
\let\uproot\relax

\usepackage{amsmath}
\usepackage{bm}
\usepackage{comment}
\usepackage{natbib}
\usepackage{graphicx}
\usepackage{xcolor}
\usepackage{ulem}
\usepackage[switch,columnwise]{lineno}

\begin{document}
\Received{}
\Accepted{}
\title{Pseudo-observation of spiral galaxies in the radio band to verify depolarization models}
\author{Yuta \textsc{Tashima}\altaffilmark{1,2}}
\altaffiltext{1}{Department of Astronomical Science, School of Physical Sciences, The Graduate University for Advanced Studies (SOKENDAI), 2-21-1 Osawa, Mitaka, Tokyo 181-8588, Japan}
\altaffiltext{2}{Department of Physics, Faculty of Sciences, Kyushu University, 744 Motooka, Nishi-ku, Fukuoka 819-0395, Japan}
\email{yuta.tashima@grad.nao.ac.jp}

\author{Takumi \textsc{Ohmura},\altaffilmark{3}}
\altaffiltext{3}{Institute for Cosmic Ray Research, The University of Tokyo, 5-1-5 Kashiwanoha, Kashiwa, Chiba 277-8582 Japan}
\email{tohmura@icrr.u-tokyo.ac.jp}

\author{Mami \textsc{Machida}\altaffilmark{4}}
\altaffiltext{4}{Division of Science, National Astronomical Observatory of Japan, 2-21-1, Osawa, Mitaka, 181-8588, Japan,[0000-0001-6353-7639]}


\maketitle

\begin{abstract}
Studies of the three-dimensional structures of galactic magnetic fields are now entering a new era, with broadband, highly sensitive radio observations and new analysis methods.
To reveal the magnetic field configuration from the observed intensities integrated along the line of sight, it is necessary to derive an appropriate model involving various combinations of parameters that can reproduce the same observational characteristics.
We aim to clarify the relationship between the radiation field and the spatial distribution of physical quantities through pseudo-observations using global three-dimensional magnetohydrodynamics (MHD) simulation results.
In particular, we focus here on using the depolarization effect, which is important in the meter-wave band, to verify the polarization model and to identify the emission region.
First, we show that wavelength-independent depolarization, which takes into account anisotropic turbulence, does not work efficiently because the polarized emission is stronger in regions of ordered spiral fields than in regions dominated by isotropic turbulent fields.
Beam depolarization, specifically internal depolarization, becomes more effective below 1$\mathrm{\> GHz}$.
Although in and close to the equatorial plane there will be strong depolarization which increases with observing wavelength, this effect is less in the halo, making halo magnetic fields detectable through their polarized emission at meter-wavelength bands.
Although polarized emission from the halo is below the detection limit of current facilities, it will be detectable within the SKA era.
In addition, we found that the spiral polarization projected on a screen is produced by overlapping magnetic flux tubes extending to different heights from the equatorial plane.
This suggests that the traditional classification of global magnetic fields has difficulty 
reproducing the global structure of the magnetic fields.
Finally, we demonstrate the method to separate magnetic flux tubes at different heights by using peak frequencies which cause the decreasing of polarized flux.
\end{abstract}


\section{Introduction}
\label{sec:intro}
Spiral galaxies typically have mean magnetic fields of a few $\mu$G, with additional turbulent magnetic fields that are stronger than the mean field \citep{soida_2011,fletcher_2010,beck_2015}.
These magnetic fields have a great influence on the time evolution of galaxies and on star-formation activity \citep{krause_2009,birnboim_2015}.
Studies of the structure, origin, and evolution of galactic magnetic fields have therefore been conducted for many years \citep[e.g.,][]{beck_2015_m,kurahara_2021}.

The polarization of the radio continuum, which originates from synchrotron radiation, is a useful tool for revealing the magnetic fields in galaxies and the interstellar medium (ISM). 
The total energy of the magnetic field, including both the turbulent field and the mean field, can be estimated from the total intensity of the radiation.
The polarized intensity provides information about the ordered field, and the strength of the magnetic field along the line of sight (LOS) can be estimated from multi-frequency observations of the Faraday rotation Measure (RM), which is the coefficient of proportionality between the polarization angle and the square of the observing wavelength \citep[e.g.,][]{rand_1989,stepanoov_2008, mao_2012}.
We can calculate the RM theoretically as the integral over the product of the electron number density $n_\mathrm{e}$ and the magnetic field parallel to the LOS $B_\parallel$.
Using the RM and the polarized intensity, the strengths of both the mean and the ordered magnetic fields can be determined.
For example, \cite{beck_2015} summarized the results of radio observations of the spiral galaxy IC\,342 and reported that it has a total field of 13 $\mu$G and an ordered field of 4 $\mu$G.
The coherence length of the turbulent magnetic field in a galactic disk is thought to be 10--100$\mathrm{\> pc}$ \citep{ohno_1993}.
The length scale of turbulent magnetic fields are reported to be about 50 pc for M51 \citep{fletcher_2011} and about 70 pc for IC 342 \citep{beck_2015}.

However, the RM sometimes cannot be interpolated linearly, especially in the low frequency region \citep[e.g.,][]{o'sullivan_2009,o'sullivan_2012}.
One of the reasons of this is that wavelength dependence of Faraday rotation leads to cancelling out of polarized emission on the long wavelength side.
The existence of turbulent magnetic fields also depolarizes emission since polarized angles are randomized by turbulent fields.
The overlaying emission region along the line of sight becomes the another reason for the decreasing of the polarized emission.
These phenomena are referred to as Faraday depolarization (see detail in section \ref{sec:dep}).
\cite{burn_1966} modeled the depolarization produced by isotropic turbulence with a Gaussian distribution.
\cite{sokoloff_1998} updated the depolarization model to treat the effect of anisotropic turbulent fields.
\cite{arshakian_2011} investigated the effect of anisotropic turbulence, and they studied the dependence of Faraday depolarization on the wavelength and magnetic field strength.
As an example of depolarization in low frequency radio observations of spiral galaxies, \cite{mulcahy_2014} reported that the 151 MHz polarized emission of M\,51 is below the detection limit of the Low Frequency Array (LOFAR).

Another important analysis method for magnetic fields is Faraday tomography, which can separate the components of the integrated intensity along the LOS in Faraday depth space \citep{brentjens_2005}.
\cite{fletcher_2011} reported that M\,51, a typical bi-symmetric spiral (BSS-type) galaxy, is divided into an axisymmetric spiral in the disk and a BSS in the halo.
In this way, Faraday tomography is expected to enable the elucidation of the magnetic field structure along the LOS.
However, the interpretation of Faraday tomography is not yet adequate because there is not a simple correspondence between Faraday depth space and real space along the LOS.

\cite{shneider_2014a} developed a theoretical model of galactic disks and halos as two-layer or three-layer magnetized plasmas.
They reported that a comparison between multi-wavelength polarization observations and two-layer or three-layer depolarization models can characterize the strength of the mean field and the anisotropy of the turbulent field in disks and halos.
Using three-band data for M51, \cite{shneider_2014b} showed that this galaxy can be explained by a two-layer model, with the polarized emission originating behind the disk being completely depolarized.
They also indicated that the turbulence characteristics of the halo are isotropic, while those of the disk are anisotropic.
\cite{kierdorf_2020} reported that even after adding S-band results to the previous data \citep{shneider_2014b}, a two-layer model remains suitable for M51. Importantly, they pointed out that the S-band results are essential for distinguishing between the two-layer and three-layer models.
This model is unique in that it includes multiple types of mean fields in the calculations.
However, these mean fields models did not consider whether the models satisfy physical constants or not.
We therefore think that it is important to investigate the depolarization characteristics using the results of numerical experiments in order to elucidate the magnetic field structure of the galaxy.

In the 2000s, several studies have begun to use magnetohydorodynamic (MHD) simulations to clarify the time evolution and origin of magnetic fields and their effects on galaxy structure formation \citep{nishikori_2006, machida_2009,machida_2013, hanasz_2009}.
\cite{machida_2013} performed a three-dimensional MHD simulation of a galactic gas disk and reported that the amplified magnetic field in the disk is constantly supplied to the halo.
Cosmological galaxy-formation simulations based on the $\Lambda$CDM model have also been performed \citep{springel_2008, pakmor_2014}.
These numerical simulations of galactic disks show that galactic magnetic fields have complex turbulent structures.

Since the physical quantities obtained in MHD simulations are different from the observed quantities,
we need to calculate the observables to understand the correspondence between three-dimensional structures of the magnetic fields in galaxies and integrated intensity map.
Some groups are starting to study pseudo-observations of radiation using the results of numerical simulations of galactic disks \citep{machida_2018,machida_2019,reissl_2019}.
Since the turbulence scale of galaxies is shorter than the numerical grid size, \cite{reissl_2019} performed pseudo-observations assuming Gaussian sub-grid turbulence fields.
They assumed the beam size of the pseudo-observations to be larger than the assumed sub-grid turbulence scale.
Therefore, the depolarization effects due to the isotropic turbulent magnetic field in the sub-grid are sufficient to affect the polarization image.
This means that depolarization strongly affects the given turbulence structure and thus that both the appropriate coherence length scale of the turbulence and the fluctuations of the magnetic field strength less than the grid size need to be considered.
\cite{machida_2018} performed pseudo-observations that take into account the wavelength-independent depolarization produced by the isotropic turbulent magnetic fields proposed by \cite{sokoloff_1998}. 
\cite{machida_2019} subsequently incorporated wavelength-dependent depolarization to determine the polarization intensity at meter wavelengths.
They reported that the magnetic field in the disk is depolarized by Faraday rotation, but that a dilute halo field may be observable at meter wavelengths.

In this paper, we present pseudo-observation results over the frequency range 0.15$\mathrm{\> GHz}$ to 8.5$\mathrm{\> GHz}$ using more accurate depolarization models.
Our purpose is to clarify the relationship between the actual structure and the polarized radiation by comparing the three-dimensional structure of the galactic magnetic field with the observed polarization.
We introduce the depolarization models in section 2 and discuss the numerical methods used to obtain the pseudo-observations in section 3.
We present numerical results in section 4 and discuss our conclusions in section 5.
Throughout this paper, we choose the LOS direction to be the $z$-axis.

\section{Depolarization}
\label{sec:dep}
The polarization degree $p$ defines the fraction of the total intensity that is polarized.
It is defined in this paper by the following equation:
\begin{equation}
    p=p_0D,
\end{equation}
where $p_0$ is the intrinsic degree of polarization of synchrotron radiation,
\begin{equation}
    p_0=\frac{3s+3}{3s+7}
\end{equation}
\citep{burn_1966}, where $s$ is the spectral index of cosmic-ray electrons, and
$D$ is the depolarization coefficient.
Since the depolarization effect depends on the configuration of the magnetic field, several examples of magnetic field structure and thermal electron number density have already been studied \citep{burn_1966, sokoloff_1998}.
When depolarization models overlap, the effect can be expressed as the product of the respective depolarization coefficients:
\begin{equation}
    D=D_WD_\mathrm{F}D_AD_B,
\end{equation}
where $D_W$ represents the effect of wavelength-independent depolarization, $D_\mathrm{F}$ represents differential Faraday rotation depolarization, and $D_A$ and $D_B$ represent two types of beam depolarization.
The following sub-sections describe these depolarization models.

\subsection{Wavelength-independent depolarization}
Since synchrotron radiation is emitted with a direction of linear polarization perpendicular to the magnetic field, the polarization angle in a turbulent magnetic field is oriented in various directions.
Therefore, the polarization of the radiation cancels out even without considering Faraday rotation.
Wavelength-independent depolarization occurs in such a turbulent magnetic field.
In this section, we assume that the synchrotron emissivity scales as $\varepsilon_I \propto B^2_{\perp}~~(s=3$, see equation \eqref{eq:em}) below.
When there is a Gaussian turbulent magnetic field with dispersion $\sigma_\mathrm{i}^2$ and average value $\overline{B}_\mathrm{i}$, the depolarization coefficient can be represented as \citep{sokoloff_1998}
\begin{equation}
    \begin{split}
        D_W=&\left(\overline{B^2}~\overline{B_\perp^2}\right)^{-1} \\
        &\times\left\{\left[\overline{B}_{\sigma\mathrm{max}}^4-\overline{B}_{\sigma\mathrm{min}}^4
        +3\left(\sigma_\mathrm{max}^4-\sigma_\mathrm{min}^4\right)\right.\right. \\
        &\left.+6\left(\overline{B}_{\sigma\mathrm{max}}^2\sigma_\mathrm{max}^2-\overline{B}_{\sigma\mathrm{min}}^2\sigma_\mathrm{min}^2\right)\right.\\ 
        &\left.+\overline{B_z^2}\left(\overline{B^2}_{\sigma\mathrm{max}}-\overline{B^2}_{\sigma\mathrm{min}}\right)\right]^2 \\
        &\left.+4\overline{B}_{\sigma\mathrm{max}}^2\overline{B}_{\sigma\mathrm{min}}^2\left[\overline{B^2}
        +2\left(\sigma_\mathrm{max}^2+\sigma_\mathrm{min}^2\right)\right]^2\right\}^{1/2} 
    \end{split}
    \label{eq:aniso}    
\end{equation}
where $\overline{B_i^2}=\overline{B}_i^2+\sigma_i^2$, and $\overline{B_\perp^2}=\overline{B}^2_\perp+\sigma_\mathrm{max}^2+\sigma_\mathrm{min}^2$.
Here, the subscript $i$ represents the direction of the magnetic field.
The quantity $\sigma_\mathrm{max}^2$ is the dispersion in the direction in which the magnetic field is most dispersed in the plane perpendicular to the LOS,
and $\sigma_\mathrm{min}^2$ is the dispersion in the direction perpendicular to direction of $\sigma_\mathrm{max}^2$.
The quantities $\overline{B}_{\sigma\mathrm{max}}$ and $\overline{B}_{\sigma\mathrm{min}}$ are the mean magnetic field strengths in each dispersion direction.
The quantities $\sigma_\mathrm{max}^2$ and $\sigma_\mathrm{min}^2$ are obtained from the dispersion in the $x$ and $y$ directions and their covariance.
We define the covariance matrix as
\begin{equation}
    \left(
    \begin{array}{cc}
        \sigma_x^2 & \sigma_{xy} \\
        \sigma_{xy} & \sigma_y^2
    \end{array}
    \right),
\end{equation}
where $\sigma_x^2$ and $\sigma_{y}^2$ are the dispersions in the $x$- and $y$-directions and $\sigma_{xy}$ is the covariance.
The eigenvalues of this matrix are $\sigma_\mathrm{max}^2$ and $\sigma_\mathrm{min}^2$, and the corresponding eigenvectors are the directions of $\sigma_\mathrm{max}^2$ and $\sigma_\mathrm{min}^2$.

In the special case where the turbulent magnetic field is isotropic ($\sigma_x=\sigma_y=\sigma_z\equiv\sigma$) and the mean field in the LOS direction is zero, equation \eqref{eq:aniso} can be rewritten as
\begin{equation}
    D_W=\frac{1+7\left(\sigma/\overline{B}_\perp\right)^2}
    {1+9\left(\sigma/\overline{B}_\perp\right)^2+10\left(\sigma/\overline{B}_\perp\right)^4}.
    \label{eq:w_iso}
\end{equation}
Previous studies \citep{machida_2018} have used equation \eqref{eq:w_iso} to calculate the wavelength-independent depolarization.

\subsection{Wavelength-dependent depolarization}

Wavelength-dependent depolarization is caused by Faraday rotation.
The amount of rotation of the plane of polarization due to Faraday rotation is
\begin{equation}
    \Delta\chi_\mathrm{F}=\mathrm{RM}\lambda^2,
    \label{eq:rm-lmd}
\end{equation}
where $\lambda$ is the observing wavelength.
We also introduce the Faraday depth (FD) as an indicator of the amount of Faraday rotation that occurs in a specific region; it is given by
\begin{equation}
     \mathrm{FD}=K\int^{z_\mathrm{a}}_{z_\mathrm{b}} n_\mathrm{e}B_zdz, 
    \label{eq:RM}
\end{equation}
\begin{equation}
    K=\frac{e^3}{2\pi m_\mathrm{e}^2c^4}=\mathrm{constant},
\end{equation}
where $n_e$ is the thermal electron number density, and $B_\parallel$ is the strength of the magnetic field parallel to the LOS.
In addition, $e$ is the elementary charge, $m_\mathrm{e}$ is the mass of the electron, and $c$ is the speed of light.
The FD matches the RM if the range of integration in equation \eqref{eq:RM} is extended to the entire LOS ($z_a\rightarrow 0, z_b\rightarrow\infty$).

\subsubsection{Differential Faraday rotation depolarization}
When plasma is composed of both thermal and cosmic-ray electrons, synchrotron radiation and Faraday rotation occur simultaneously at the same position.
Therefore, the polarization angle inside that region becomes irregular, and the polarized intensity averaged over that region decreases even if the polarization angle emitted by the synchrotron radiation is uniform along the LOS.
This phenomenon is called differential Faraday rotation depolarization.
The depolarization coefficient can be written as
\begin{equation}
    D_\mathrm{F}=\frac{\sin(\mathrm{FD}\lambda^2)}{\mathrm{FD}\lambda^2},
    \label{eq:dif}
\end{equation}
in the region where $n_\mathrm{e}$ and magnetic fields $\bm{B}=(B_x,B_y,B_z)$ are constants.

\subsubsection{Beam depolarization}
\label{sec:beam}
When polarized radiation passes through a turbulent magnetic field, the polarization angle becomes randomized due to Faraday rotation.
These random polarization angles all enter the beam and cancel each other out; this is called \lq\lq beam depolarization.\rq\rq\  

Beam depolarization can be divided into two types: one is internal depolarization that affects the radiating region and the other is external depolarization caused by the foreground medium.
We do not consider external depolarization in this study.

When the emission of polarized radiation and Faraday rotation occur at the same position, internal depolarization becomes important.
Consider the case of Gaussian turbulent magnetic field with the dispersion  $\sigma_z^2$ and constant polarized radiation between the path lengths $L$ and $L+\Delta z$. 
The degree of polarization observed along this LOS is 
\begin{equation}
    \begin{split}
        p =& p_0 \exp\left\{-2(Kn_\mathrm{e}\sigma_z)^2L\lambda^4d\right\}\\
        &\times \left(\frac{1-e^{-S'}}{S'}\right),\\
        S' =&\left(Kn_\mathrm{e}\sigma_z\right)^2d\Delta z\lambda^4,
    \end{split}
    \label{eq:dp_in_m}
\end{equation}
where $d$ is the coherence length scale of the turbulent field.
Equation \eqref{eq:dp_in_m} is the product of depolarization in the local radiation area (between $L$ and $L+\Delta z$) and depolarization between the local radiation area and the observer.
We call the former "internal depolarization A" and the latter "internal depolarization B".
The respective depolarization coefficients are
\begin{equation}
    D_A=\left|\frac{1-e^{-S'}}{S'}\right|,
    \label{eq:DA}
\end{equation}
\begin{equation}
\begin{split}
    D_B&=\exp\left\{-2(Kn_\mathrm{e}\sigma_z)^2dL\lambda^4\right\}\\
    &=\exp\left\{-2\sigma^2_{\mathrm{FD}}\frac{L}{d}\lambda^4\right\},
    \label{eq:DB}
\end{split}
\end{equation}
where $\sigma^2_\mathrm{FD} \equiv (Kn_\mathrm{e}\sigma_zd)^2$ is the dispersion of the FD.

\section{Pseudo-observations}

\subsection{Polarized radiative transfer equation}

We first present the equations of radiative transfer that include depolarization.
Here, we use the Cartesian coordinate $S$-frame $(x,y,z)$ in which the $z$-axis is the LOS direction of the radiation:
\begin{align}
        \frac{dI}{dz}&=\varepsilon_I\\
        \frac{dQ}{dz}&=D\varepsilon_Q\cos2\chi_B-2U\frac{d\chi_\mathrm{F}}{dz}\\     
        \frac{dU}{dz}&=D\varepsilon_Q\sin2\chi_B+2Q\frac{d\chi_\mathrm{F}}{dz},    
    \label{eq:rt}
\end{align}
where $(I,Q,U)$ are the Stokes parameters, and $\varepsilon_{I, Q}$ are the radiation coefficients of each Stokes parameters. 
The quantity $d\chi_\mathrm{F}/dz~~(\equiv Kn_\mathrm{e}B_z\lambda^2)$ is the Faraday rotation term,
$D$ is the depolarization coefficient described in section \ref{sec:dep}, and
$\chi_B$ is the direction of $B_\perp$.
The transfer equations can be solved easily by taking the direction of the magnetic field to be the direction of Stokes $U$.
In this study, we confirm that synchrotron self-absorption and free-free absorption are not effective, even at 0.1 GHz.

The polarization angle $\chi$ and the polarization intensity is calculated as follows using the Stokes parameters:
\begin{equation}
    \chi = \left\{
    \begin{array}{ll}
    \frac{1}{2} \tan^{-1}{\left(\frac{U}{Q}\right)}&\quad(Q>0), \\
    \frac{1}{2} \tan^{-1}{\left(\frac{U}{Q}\right)}+\frac{\pi}{2}&\quad(U>0, Q<0),\\
    \frac{1}{2} \tan^{-1}{\left(\frac{U}{Q}\right)}-\frac{\pi}{2}&\quad(U<0, Q<0),\\
    \frac{\pi}{4}&\quad(U>0, Q=0),\\
    -\frac{\pi}{4}&\quad(U<0, Q=0),\\
    0&\quad(U=0,Q=0),\\
    \end{array}
\right.
\end{equation}
\begin{equation}
    P = \sqrt{Q^2 + U^2}.
\end{equation}

The number density of cosmic-ray electrons is required to calculate synchrotron radiation.
However, the cosmic-ray electron distribution cannot be obtained from the MHD simulation.
Consequently, we assume that the energy density of cosmic-ray particles is equal to the magnetic energy density \citep{akahori_2018}.
Also, we assume  a single power-law distribution for the cosmic-ray electrons.
Thus, the number density of cosmic-ray particles between Lorentz factors $\gamma$ and $\gamma+d\gamma$ is given by 
\begin{equation}
    n_\mathrm{ce}(\gamma)=N_0\gamma^{-s}d\gamma,
    \label{eq:cosmicray}
\end{equation}
where $s$ is the spectral index; in this study, $s=3$ \citep{sun_2008}. 
Therefore, the cosmic-ray electron number density $n_{ce}$ is
\begin{equation}
    \begin{split}
        n_\mathrm{ce}&=N_0\int^{\gamma_\mathrm{max}}_{\gamma_\mathrm{min}}\gamma^{-s}d\gamma\\
        &=N_0N_1(s,\gamma_\mathrm{min},\gamma_\mathrm{max}),
    \end{split}
\end{equation}
where $N_1=[\gamma^{1-s}]^{\gamma_\mathrm{max}}_{\gamma_\mathrm{min}}/(1-s)$.
Similarly, the energy density is
\begin{equation}
    \begin{split}
        \epsilon_\mathrm{ce}&=N_0m_\mathrm{e}c^2\int^{\gamma_\mathrm{max}}_{\gamma_\mathrm{min}}\gamma^{1-s}d\gamma\\
        &=N_0m_\mathrm{e}c^2N_2(s,\gamma_\mathrm{min},\gamma_\mathrm{max}),
    \end{split}
\end{equation}
where $N_2=[\gamma^{2-s}]^{\gamma_\mathrm{max}}_{\gamma_\mathrm{min}}/(2-s)$, and
$\gamma_\mathrm{min}$ and $\gamma_\mathrm{max}$ are free parameters.
In this calculation, we used $\gamma_\mathrm{min}=200$, $\gamma_\mathrm{max}=3000$.
The quantities $n_\mathrm{ce}$ and $\epsilon_\mathrm{ce}$ are almost independent of the value of $s$ when $s=3$.
Additionally, the selection of $\gamma_\mathrm{max}$---anywhere from 3000 to infinity---has little effect on radio emission in the GHz band.
On the low-energy side ($\gamma<200$), the cosmic-ray electrons lose energy by Coulomb interactions with the thermal plasma \citep{sarazin_1999}. 

We assume an equipartition between the total energy in relativistic particles, which is twice the energy in relativistic electrons, and the magnetic energy:
\begin{equation}
    2\epsilon_\mathrm{ce}=\frac{B^2}{8\pi} .
\end{equation}
Therefore,
\begin{equation}
    N_0=\frac{B^2}{16\pi m_\mathrm{e}c^2N_2},
    \label{eq:n_0}
\end{equation}
and the cosmic-ray electron number density is
\begin{equation}
    n_\mathrm{ce}=\frac{N_1B^2}{16\pi m_\mathrm{e}c^2N_2}.
\end{equation}
Using $N_0$ obtained from Equation \eqref{eq:n_0}, we obtain the radiation coefficients 
\begin{equation}
    \varepsilon_I=c_5(s)N_0B_\perp^\frac{s+1}{2}\left(\frac{\nu}{2c_1}\right)^\frac{1-s}{2},\quad
    \varepsilon_Q=\frac{s+1}{s+7/3}\varepsilon_I,
    \label{eq:em}
\end{equation}
where
\begin{equation}
    \begin{split}
        c_1&=\frac{3e}{4\pi m_\mathrm{e}^3c^5}, \\
        c_5(s)&=\frac{\sqrt{3}}{16\pi}\frac{e^3}{m_\mathrm{e}c^2}\left(\frac{s+7/3}{s+1}\right)\Gamma\left(\frac{3s-1}{12}\right)\Gamma\left(\frac{3s+17}{12}\right), \\
    \end{split}
\end{equation}
The quantities $c_1$ and $c_5$ are dimensionless functions of $s$, which have been determined by \cite{pacholczyk_1970}.

\subsection{The galaxy model}
\subsubsection{3D MHD simulation}
\begin{table}
\caption{Numerical units of the physical quantities from the MHD simulation used for the pseudo-observations}
\begin{tabular}{lc}
\hline
\hline
Physical Quantity & Numerical Unit\\
\hline
Density & $1.6\times10^{-24}$ $\mathrm{g~cm^{-3}}$\\
Magnetic Field & 26 $\mathrm{\mu G}$\\
Temperature & $5.2\times10^6$ K\\
Cosmic-ray electron number density & $3.5\times10^{-3}$ $\mathrm{cm^{-3}}$\\
Absorption coefficient (0.1$\mathrm{\> GHz}$) & $6.7\times10^{-16}$\\
\hline
\end{tabular}
\label{table:1}
\end{table}

In this paper, we calculate the observables using the results of MHD simulations \citep{machida_2013} that were performed in the cylindrical coordinate system $(r, \phi, z')$.
The size of the simulation box is $r$ (kpc) $< 56$ in the radial direction, $0 \leq \phi \leq 2\pi$ in the azimuthal direction, and $-10 <z'$ (kpc) $<10$ in the vertical direction.
The numbers of grid points are (250,256,640) in $(r, \phi, z')$ direction, respectively.
For $0 < r$ (kpc) $< 6$ and $-2 < z'$ (kpc) $< 2$---near the center of the galaxy model---the spatial resolution of the simulation is $\Delta r = 50\mathrm{\> pc}$ and $\Delta z'= 10\mathrm{\> pc}$.
The azimuthal resolution is 25$\mathrm{\> pc}$ at $r = 1\mathrm{\> kpc}$.
The time resolution of the simulation at this spatial resolution is $10^3$ yr.
The growth timescale of the magnetorotational instability (MRI) is comparable to the rotation time. The rotation time at 1 kpc is about $10^4$ yr, while our numerical simulation extends over $5.8\times 10^9$ yr; in other words, this time is long enough to resolve the MRI growth time.

The initial condition of the simulation was an equilibrium torus threaded by a weak azimuthal magnetic field embedded in a non-rotating, high-temperature corona under Miyamoto and Nagai potentials \citep{miyamoto_1975}.
See \cite{machida_2013} for more details of the simulation data.
Since \cite{machida_2013} is an ideal MHD simulation, it is a scale-free simulation.
For this study, we therefore chose parameters close to those appropriate for a spiral galaxy, as shown in Table \ref{table:1}.
The cosmic-ray electron number density and the radiation coefficients required for the radiative transfer equation are determined automatically from equations \eqref{eq:n_0} and \eqref{eq:em} using the physical quantities obtained from the MHD simulation (see also Table \ref{table:1})

\subsubsection{Coordinates}
\begin{figure}
    \begin{center}
    \includegraphics[width=0.8\linewidth]{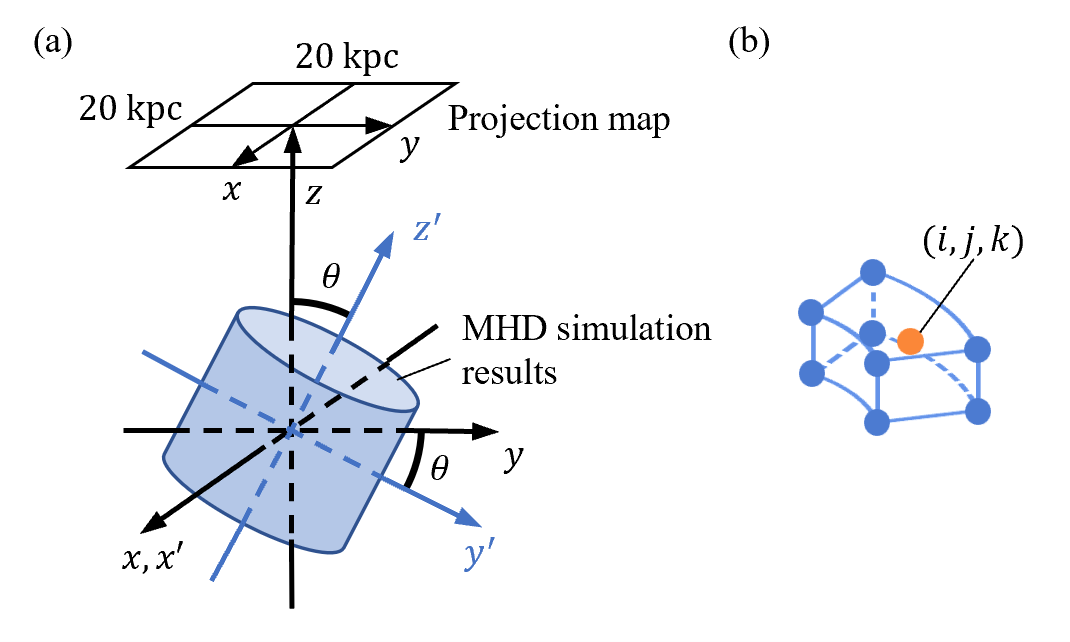}
    \end{center}
    \caption{
    (a) The $z$-axis is defined to be the LOS direction, which is normal to the sky projection plane $(x, y)$.
    The primed axes indicate the original coordinates of the galaxy in the numerical simulation.
    The angle theta is the inclination angle between the rotational axis of the galaxy (the $z'$-axis) and the LOS.
    (b) To determine the depolarization coefficient, we need the dispersion for each magnetic field direction at point $(x,y,z)$ in the S-frame, which is denoted by the orange dot in this figure.
    In this work, the dispersions are computed from equation \eqref{eq:sigma} by using the magnetic fields at the nearest eight grids in the $S'$-frame, which are denoted by the blue dots. 
    }
    \label{fig:los}
\end{figure}

We calculated the observational quantities---such as the Stokes parameters and the RM---from the numerical simulation results for the galactic gaseous disk.
We set the projection plane for the pseudo-observables and its normal to be the $S$-frame $(x, y, z)$.
The $S'$-frame indicates the coordinates for the model galaxy obtained from the three-dimensional global simulation. 
Note that the center of the $S$-frame is at the same location as that of the $S'$-frame and that the $x$-axis overlays the $x'$-axis (see figure \ref{fig:los}a).
The MHD variables in the $S$-frame are calculated by linear interpolation of the variables in the S$'$-frame.
The inclination angle between the rotation axis of the galaxy (the $z'$-axis) and the LOS (the $z$-axis) is defined as $\theta$.
Face-on and edge-on correspond to $\theta=0^\circ$ and $90^\circ$, respectively. 
In this study, we report the results for the face-on view ($\theta=0^\circ$).
The spatial range of the pseudo-observation coordinates is set to be $(x,y,z)\in\pm10\mathrm{\> kpc}$ centered on the center of the galaxy, and the numbers of grid points are $(n_x,n_y, n_z)=(200,200,200)$.
An observational map can be obtained by numerical integration from $z=-10\mathrm{\> kpc}$ to $z=10\mathrm{\> kpc}$ at all 2D coordinate points $(x,y)$.
Therefore, a pseudo-observation result can be obtained as an image with the numbers of pixels $(N_z, N_y)=(200,200)$ centered on the galactic center.

\subsubsection{Determination of the magnetic field dispersion}

It is necessary to take into account the depolarization induced by turbulent fields with length scales below the grid size.
In this study, we implement the depolarization effect of the turbulent field by using the model introduced in section \ref{sec:dep}.

To use this depolarization model, we need to obtain the mean field and the sub-grid-scale turbulence field.
In this study, we estimate the property of the magnetic field inside the grid from the surrounding magnetic fields, following \cite{machida_2019}.
The validity of this method is described in section 4.1 of \cite{machida_2019}.
First, we calculate the mean magnetic field by averaging the magnetic field inside the $\pm 5$ grid points in the $S'$-frame.
The mean magnetic field in the S-frame is calculated by linear interpolation using the mean field of the S-frame.
To calculate the dispersion for each magnetic field direction at the grid point $(x,y,z)$ in the S-frame, we use the eight grid points in the $S'$-frame (h), that are nearest to the given grid point, as shown in figure \ref{fig:los}b.
The dispersion of the magnetic field in the $x$-direction is thus given by:
\begin{equation}
    \sigma^2_{x}(x,y,z) =\frac{1}{8}\sum_{(l,m,n) \in h}\left[B_{x}(l,m,n) - \overline{B_{x}}(l,m,n)\right)^2.
    \label{eq:sigma}
\end{equation}
Dispersions in the $y$ and $z$ directions are obtained in the same way.
The FD dispersion $\sigma^2_\mathrm{FD}$ is also needed in order to use the internal depolarization model [see equation \eqref{eq:DB}).
The dispersion of FD in grid point $(x,y,z)$ is
\begin{equation}
\begin{split}
    \sigma^2_\mathrm{FD}(x,y,z)     
    =\frac{1}{9} 
    \sum^{x+1}_{x''=x-1} \sum^{y+1}_{y''=y-1}&\left\{\mathrm{FD}_{\mathrm{turb}}(x'',y'',z)\right.\\
    &\left.-\overline{\mathrm{FD}}_{\mathrm{turb}}(x,y,z)\right\}^2,
    \label{eq:sigma_FD}
\end{split}
\end{equation}
\begin{equation}
\begin{split}
    \mathrm{FD}_\mathrm{turb}(x,y,z)=&0.81\sum_{z'=1}^z n_\mathrm{e}(x,y,z')\\
    &\times\left(B_z(x,y,z')-\overline{B_z}(x,y,z')\right)\Delta z,   
\end{split}
\end{equation}
\begin{equation}
    \overline{\mathrm{FD}}_\mathrm{turb}(x,y,z)
    =\frac{1}{9} 
    \sum^{x+1}_{x''=x-1} \sum^{y+1}_{y''=y-1}\mathrm{FD}_\mathrm{turb}(x'',y'',z),
\end{equation}
where $\mathrm{FD}_{\rm turb}$ is the value of FD due to the sub-grid-scale turbulence and $\overline{\mathrm{FD}}_{\rm turb}$ is the average of $\mathrm{FD}_{\rm turb}$.
In this study, we assumed the length scale of the magnetic field [see equations \eqref{eq:DA} and \eqref{eq:DB}] to be $d = 100\mathrm{\> pc}$, which is the same as the length scale in a previous study \citep{machida_2019}.

\section{Results}
The pseudo-observations are performed using a dataset similar to that of \cite{machida_2018}.
We first present the RM maps and then show the results of the depolarization behavior in the radio band.

\subsection{Comparison of two methods for calculating the Faraday RM}
\label{sec:rm}

\begin{figure}
\begin{center}
\includegraphics[width = 0.9\linewidth]{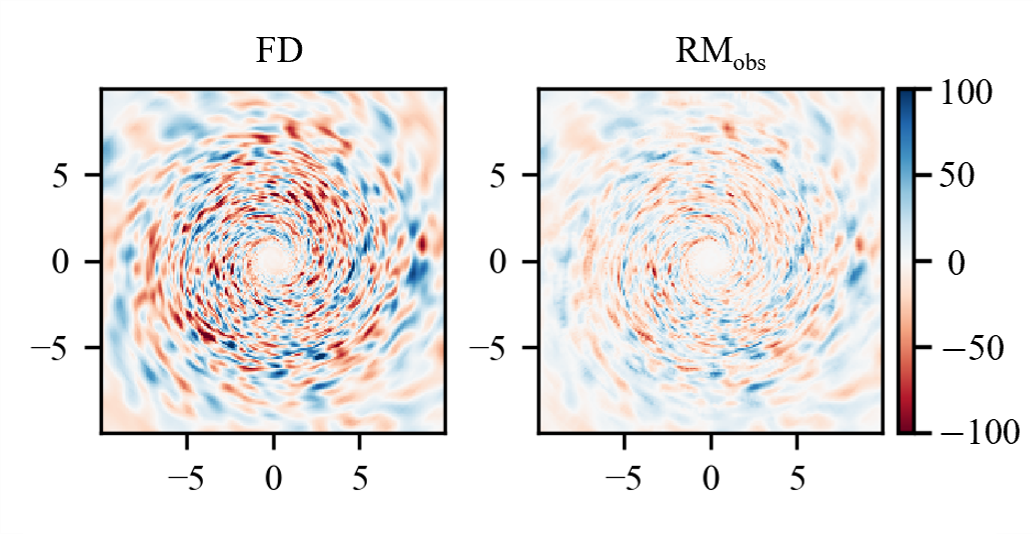}
\end{center}
\caption{
    (left) Face-on view color map of the FD of the spiral galaxy.
   The FD is derived using equation \eqref{eq:RM}, and it has units of [$\mathrm{rad\>m^{-2}}$].
    The vertical and horizontal axes have units of [kpc].
    (right) The RM color map calculated from equation \eqref{eq:rm-lmd2}, obtained using 8$\mathrm{\> GHz}$ and 10$\mathrm{\> GHz}$.
    }
\label{fig:RM}
\end{figure}

As mentioned in Section \ref{sec:intro}, the RM can be calculated using either the theoretical relationship (equation \eqref{eq:RM}) and the conventional relationship (equation \eqref{eq:rm-lmd}).
In this paper, we call the former results FD and the latter ${\rm RM_{obs}}$.

\subsubsection {Faraday depth}
The left panel of figure \ref{fig:RM} shows the spatial distribution of FD, which is the same result as in a previous study [figure 2a of \citep{machida_2018}].
The length scale of the galactic magnetic field in the azimuthal direction is longer than that of the vertical field, so the FD distribution shows a mottled structure extending in the azimuthal direction.
Moreover, since the gas density at the outer edge of the disk ($> 5 \mathrm{\> kpc}$) is low, the absolute value of FD is less than half that at the inner edge.

\subsubsection{Faraday RM}
The right panel of Figure \ref{fig:RM} shows the distribution of ${\rm RM_{obs}}$ obtained using the observing frequencies 8$\mathrm{\> GHz}$ and 10$\mathrm{\> GHz}$.
\begin{equation}
    \label{eq:rm-lmd2}
    {\rm RM_{obs}} = \frac{\chi_{\rm 10 GHz} - \chi_{\rm 8 GHz}}{\lambda^2_{\rm 10 GHz} - \lambda^2_{\rm 8 GHz}}.
\end{equation}
In this frequency range, the amount of Faraday rotation is small.
Therefore, the n-$\pi$ ambiguity do not appear on any pixel on the observation map.

Compared with FD, the absolute value of ${\rm RM_{obs}}$ is reduced overall.
This is consistent with the theory that the observed RM is halved when the magnetic field is symmetric along the LOS \citep{sokoloff_1998}.
On average, the absolute value of ${\rm RM_{obs}}$ is about half the value of FD, although there are some areas where it is reduced to 1/4.
In addition, the overall structure is almost the same, although if we examine the details, we can see that the sign is reversed and that the absolute value is not simply smaller.

\subsection{Effect of depolarization}


\begin{figure}
\begin{center}
\includegraphics[width=0.8\linewidth]{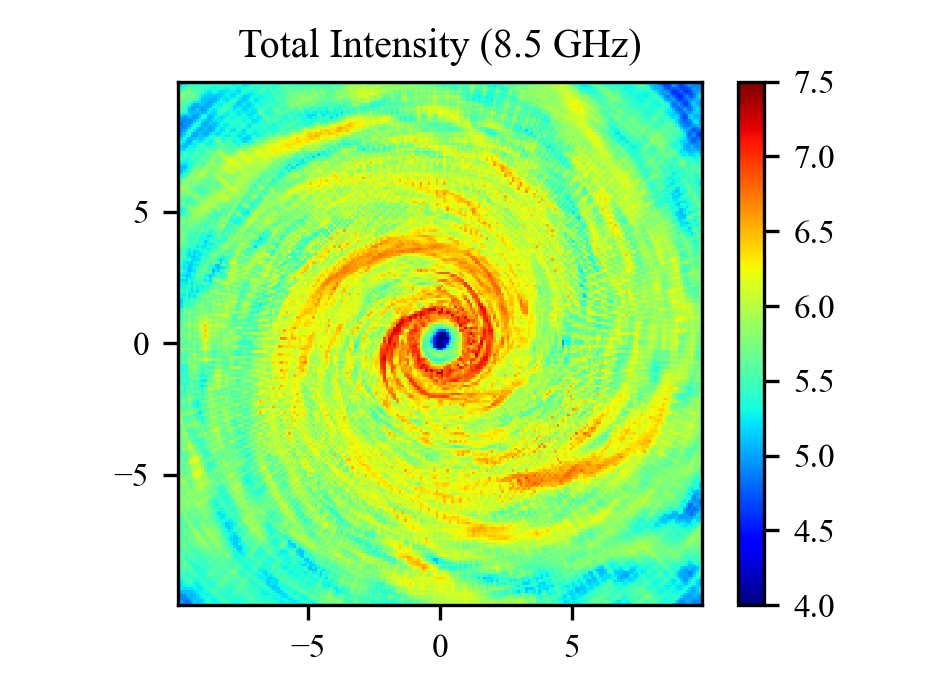}
\end{center}
\caption{
    Total intensity map of a spiral galaxy for the viewing angle $\theta=0^\circ$.
    This is the result at the pseudo-observation frequency of 8.5$\mathrm{\> GHz}$.
    The vertical and horizontal axes have units of [kpc], and the total intensity has units of [$\mathrm{mJy\>str^{-1}}$]
    }
\label{fig:si}
\end{figure}

In this section, we focus on the effect of depolarization.
First, we show the total intensity map at 8.5 GHz for comparison (Figure \ref{fig:si}).
This map reflects the magnetic spiral arm structure of the galactic disk and is consistent with our previous study \citep{machida_2018}.

\subsubsection{Wavelength-independent depolarization}
\label{sec:w}

\begin{figure}
    \begin{center}
    \includegraphics[width=1.0\linewidth]{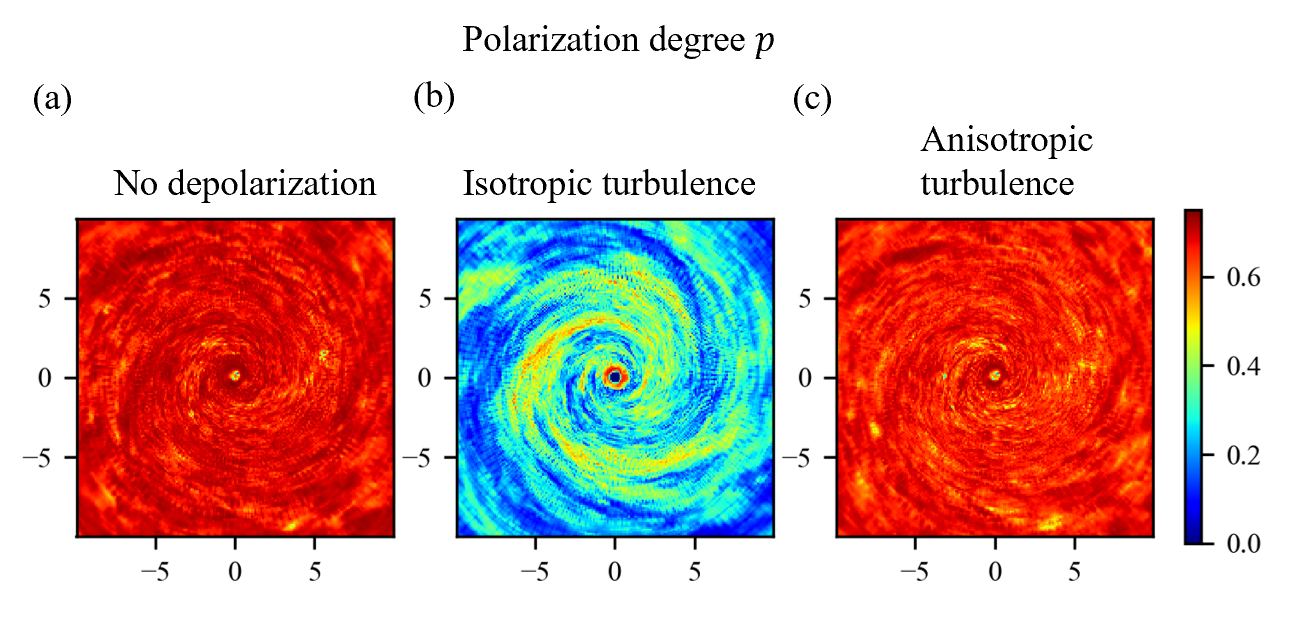}
    \end{center}
    \caption{
    These maps show the degree of polarization of a face-on spiral galaxy at the observing frequency 8.5$\mathrm{\> GHz}$.
    (a) Map of the degree of polarization without including depolarization models.
    (b) Wavelength-independent depolarization model with isotropic turbulent magnetic fields (equation \ref{eq:w_iso}).
    (c) Wavelength-independent depolarization model with anisotropic turbulent magnetic fields (equation \ref{eq:aniso}).
    }
    \label{fig:dp_w}
\end{figure}

\begin{figure}
    \begin{center}
    \includegraphics[width=0.8\linewidth]{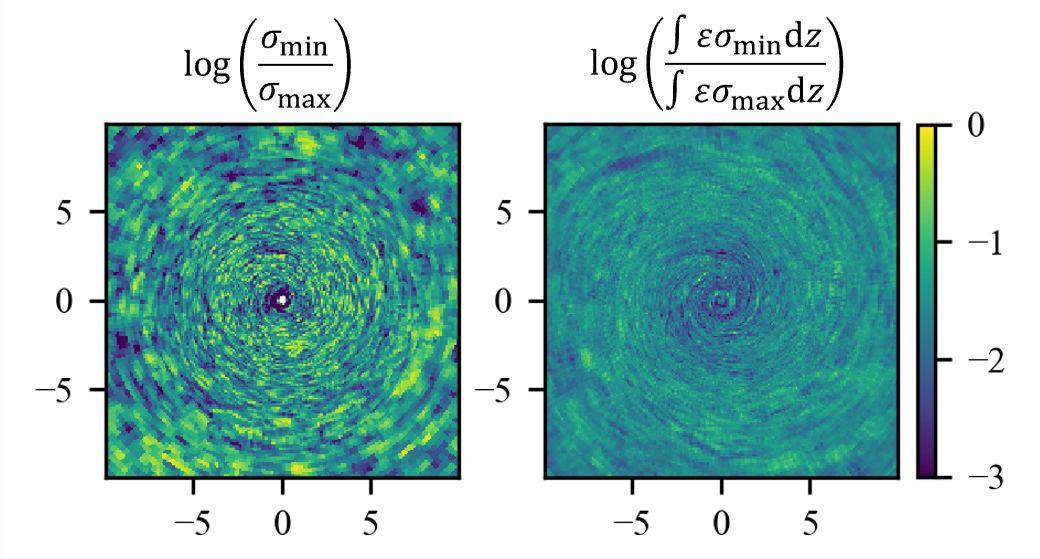}
    \end{center}
    \caption{
    (left) The ratio of $\sigma_\mathrm{max}^2$ to $\sigma_\mathrm{min}^2$.
    This panel shows the value in the equatorial plane ($z=0$).
    (right) The ratio of the values integrated along the LOS by weighting each dispersion ($\sigma_\mathrm{max}^2$ and $\sigma_\mathrm{min}^2$) with the emissivity.
    }
    \label{fig:sigma}
\end{figure}

Figure \ref{fig:dp_w} shows the effects of wavelength-independent depolarization.
Figures \ref{fig:dp_w}a, \ref{fig:dp_w}b, and \ref{fig:dp_w}c show the case with no depolarization, with isotropic turbulence, and with anisotropic turbulence, respectively.
The degree of polarization reaches 0.7 for the anisotropic model, which is consistent with almost fully polarized radiation.
When we adopt an isotropic turbulent field in the calculation, the depolarization works effectively, and the results are similar to previous studies.

The azimuthal component of the galactic magnetic field is the most substantial because of galactic rotation.
Since the energy of the turbulent magnetic field is calculated assuming a Gaussian distribution, the azimuthal turbulent magnetic field is also the dominant component.
In figure \ref{fig:dp_w}c, the wavelength depolarization is calculated taking into account anisotropic turbulence.
In this case, the degree of polarization exceeds 0.6 over the entire region due to the weakened effect of depolarization.

Next, we consider how and where anisotropic turbulence affects inefficient depolarization.
Figure \ref{fig:sigma} shows the ratio of the dispersion in the maximum and minimum directions. The local values on the equatorial plane are shown in the left panel, and the ratio of the emissivity-weighted dispersion integrated along the LOS is plotted in the right panel.

When the turbulence is completely isotropic, the dispersion ratio $\sigma_{\rm min}^2/\sigma_{\rm max}^2$ is 1.
Since the mean fields have a predominantly azimuthal component and form filamentary structures, the dispersion ratios in such filaments show anisotropic features.
Other regions where isotropic turbulence develops have $\sigma_{\rm min}^2/\sigma_{\rm max}^2 \sim 1$.
As the volume of the filamentary region is sufficiently small, the structure of the local dispersion ratio becomes sub-isotropic (figure \ref{fig:sigma}, left). 
In the presence of strongly polarized radiation from a filament, only the dispersion around the radiating region needs to be considered, since depolarization due to the turbulent field only acts on such a filamentary radiating region.
The right side of figure \ref{fig:sigma} shows the ratio of the integrals of the minimum and maximum dispersions weighted by the emissivity.
As a result, the effect of wavelength-independent depolarization is suppressed due to the anisotropy of the dispersion and of the radiating regions.
In addition, when magnetic turbulence is superimposed on the mean field, the effect of depolarization is overestimated by assuming isotropic turbulence.

\subsubsection{Differential Faraday rotation depolarization}
\begin{figure}
    \begin{center}
        \includegraphics[width=0.9\linewidth]{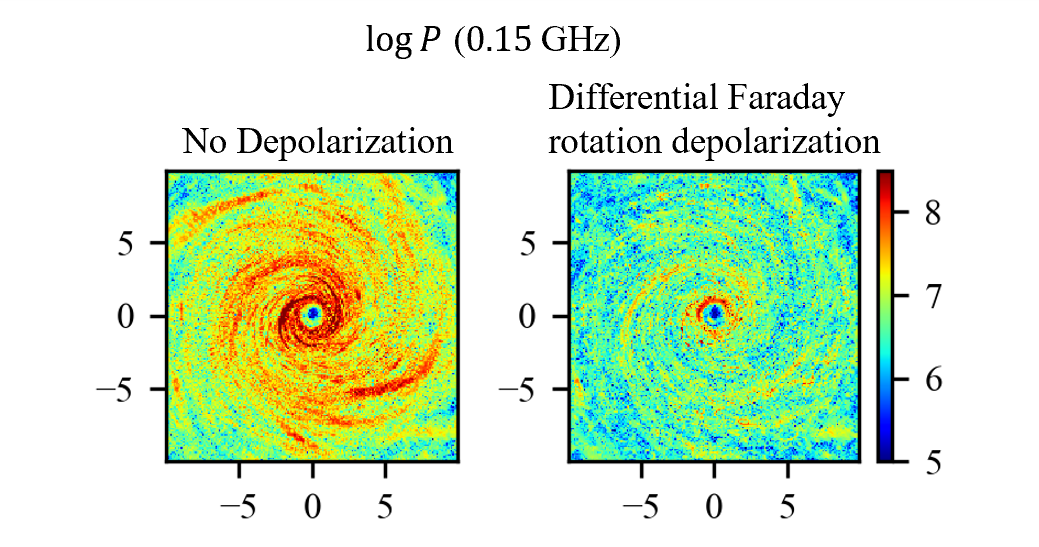}
    \end{center}
    \caption{
    (left) Map of polarized intensity at the observing frequency 0.15$\mathrm{\> GHz}$. without including depolarization models.
    The polarized intensity has units of [$\mathrm{mJy\>str^{-1}}$].
    (right) Same as (left), but including the model for differential Faraday rotation depolarization given by equation \eqref{eq:dif}.
    }
    \label{fig:sp_dif}
 \end{figure}

Figure \ref{fig:sp_dif} shows the polarization intensity at 0.15$\mathrm{\> GHz}$ for the no depolarization model (left) and including differential Faraday rotation depolarization (right), respectively.
The distribution is almost identical to that in the left panel of figure \ref{fig:sp_dif} and in figure \ref{fig:si} because depolarization is ignored, although the polarization map does highlight the spiral arms.
When differential polarization due to Faraday rotation is included in the calculation, the polarized intensity is reduced by more than two orders of magnitude compared to the unpolarized case (see the right panel of figure \ref{fig:sp_dif}).
Although the magnetic spiral arms emit strongly polarized radiation, they are effectively depolarized due to Faraday rotation. Therefore, the spiral arm region and the inter-arm region emit similar amounts of polarized radiation.
As a result, the polarization map that includes depolarization has a smaller range of intensities than that without depolarization.
In particular, the filamentary structure in the range $-8<x<-3$, $y=7.5$ is completely depolarized and invisible.
This occurs because differential Faraday rotation depolarization is more effective when the RM is larger.

\subsubsection{Beam depolarization}

\begin{figure}
    \begin{center}
    \includegraphics[width=1.0\linewidth]{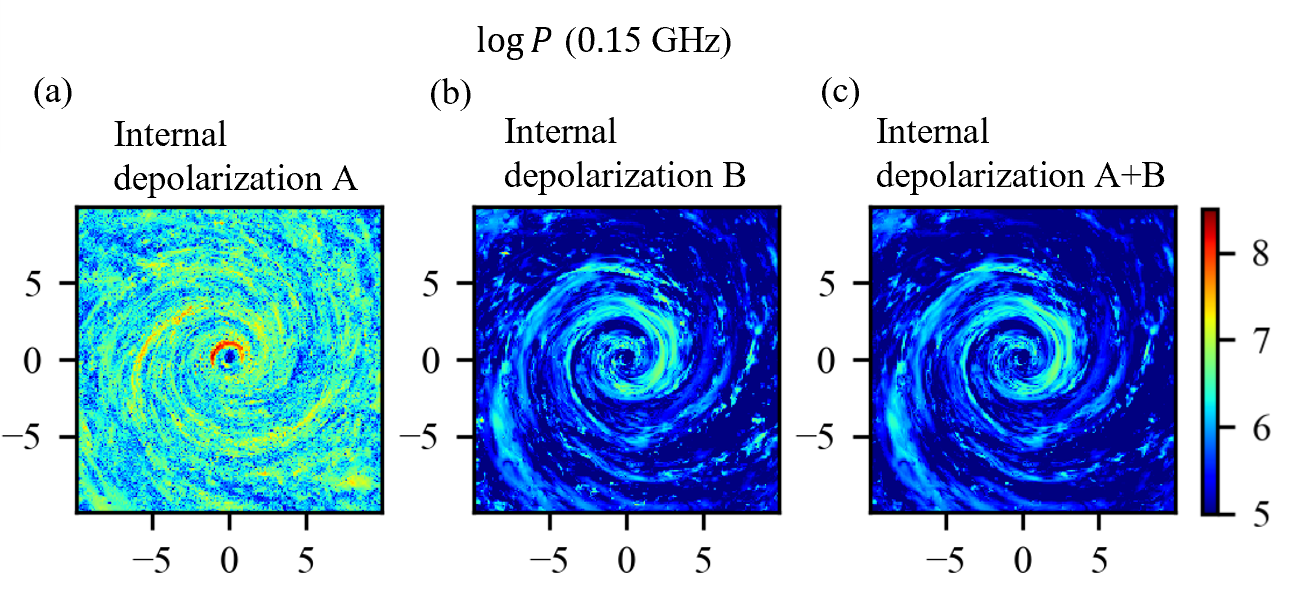}
    \end{center}
    \caption{
    (a) Maps of the polarized intensity for a face-on spiral galaxy at the observing frequency 0.15$\mathrm{\> GHz}$.
    The polarized intensity has units of [$\mathrm{mJy\>{str}^{-1}}$].
    This map includes only internal depolarization A derived from equation \eqref{eq:DA}.
    (b) Same as (a), but the depolarization model includes only internal depolarization B derived from equation \eqref{eq:DB}.
    (c) Same as (a), but including both internal depolarization A and B.
    }
    \label{fig:dp_in}
\end{figure}

Figure \ref{fig:dp_in} shows the effect of internal depolarization that is caused by the dispersion of the magnetic field along the LOS.
As shown in section \ref{sec:beam}, the internal depolarization is divided into two parts when the emitting source itself becomes the source of depolarization.
One is internal depolarization A, which originates in the local emitting region, and the other is internal depolarization B, that occurs when the polarized radiation passes through the region causing Faraday rotation in the polarization source.

Figure \ref{fig:dp_in}a shows a map of the polarized intensity that includes only internal depolarization A caused by the turbulent magnetic field.
The remaining polarized intensity due to internal depolarization A has a distribution similar to that of the total intensity map (see figure \ref{fig:si}). 
This means that the strong-field region is hardly affected by internal depolarization A.
The cause of internal depolarization A is the turbulent magnetic field in the emitting area.
Therefore, this depolarization effect is weakened in a region with ordered magnetic flux.
Since the polarized intensity averaged along the magnetic spiral arms is about $10^{7}\>\mathrm{mJy\>str^{-1}}$, internal depolarization A is more efficient than differential Faraday rotation depolarization (see Figure \ref{fig:sp_dif}, right).

Next, we consider internal depolarization B.
This corresponds to depolarization along the LOS between the emitting region and the surface of the polarized-light source.
The dispersion that causes this depolarization is therefore evaluated by integrating along the LOS where the radio waves pass.
Since the turbulent magnetic field is distributed over the galactic disk and halo, internal depolarization B works very efficiently indeed (Figure \ref{fig:dp_in}b).
The region with the remaining polarized emission is not very consistent with the total intensity map (figure \ref{fig:si}).
Figure \ref{fig:dp_in}c shows the polarized intensity map including both internal depolarizations A and B.
The overall structure is similar to the result for internal polarization B.

\subsubsection{All depolarization models}

\begin{figure}
    \begin{center}
    \includegraphics[width=0.9\linewidth]{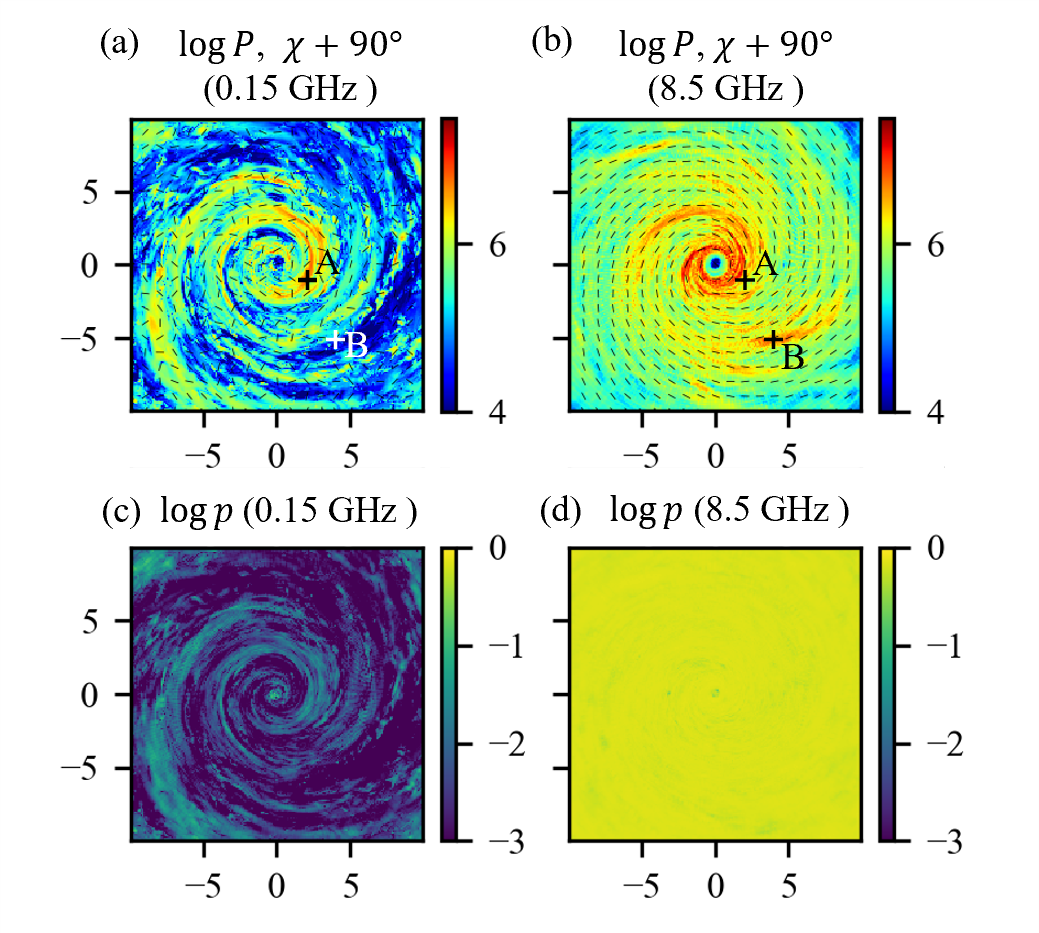}
    \end{center}
    \caption{
    (a) Map of the polarized intensity at the observing frequency 0.15$\mathrm{\> GHz}$.
    All the depolarization models introduced in section \ref{sec:dep} are considered.
    The polarized intensity has units of [$\mathrm{mJy\>str^{-1}}$].
    The observed magnetic field directions, $\chi+90^\circ$, at the same frequency are drawn on the map.
    (b) Same as (a), but the observing frequency is 8.5$\mathrm{\> GHz}$.
    (c) Map of the degree of polarization at the observing frequency 0.15$\mathrm{\> GHz}$.
    (d) Same as (c), but at the observing frequency 8.5$\mathrm{\> GHz}$.
    The crosses labeled A and B in panels (a) and (b) are the two lines of sight discussed in Section 5.2.
    }
    \label{fig:sp_1_10}
\end{figure}

Figure \ref{fig:sp_1_10} shows the polarized intensity and the degree of polarization at the observing frequencies 0.15$\mathrm{\> GHz}$ and 8.5$\mathrm{\> GHz}$.
At 8.5$\mathrm{\> GHz}$, the polarized intensity map shows a distribution similar to the total intensity map (figure \ref{fig:si}), and the degree of polarization shows high values throughout the map.
In addition, the polarization angle $\chi+90^\circ$ indicates the direction of the magnetic field aligned with the spiral structure.
This represents the direction of the disk magnetic field.
The pseudo-observation result at 0.15$\mathrm{\> GHz}$ shows that the polarized intensity is significantly reduced by the depolarization models; the polarization degree is less than 1\% almost everywhere in the map.
The polarized intensity map does not simply decrease uniformly but appears to reflect radiation from different structures.
The polarization angle completely loses information about the original magnetic field direction because the very large Faraday rotation effect makes it random.

\section{Discussion}
\subsection{The relation between FD and RM}

\begin{figure}
\begin{center}
\includegraphics[width=0.8\linewidth]{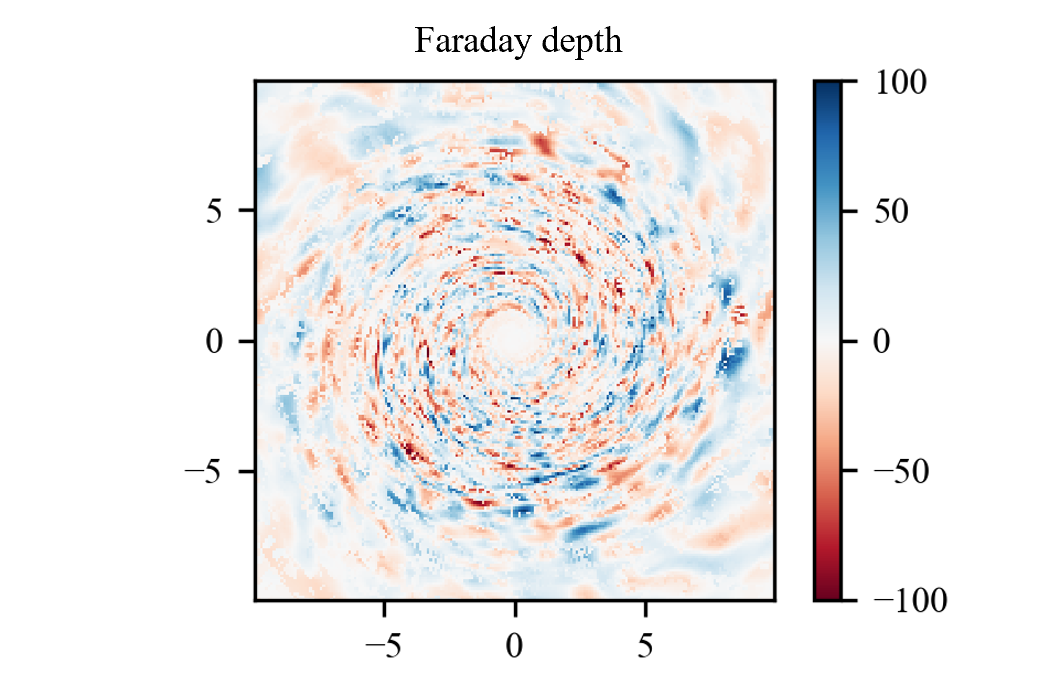}
\end{center}
\caption{
    The same as figure \ref{fig:RM}, but with the FD derived by integrating from the position of the strongest magnetic field along the LOS.
    }
\label{fig:rm3}
\end{figure}

In section \ref{sec:rm}, we compared the quantity FD calculated using equation \eqref{eq:RM} with the quantity $\rm RM_{obs}$ obtained from the proportionality relationship between the polarization angle and the square of the wavelength [equation \eqref{eq:rm-lmd}].
We found that FD is about twice as large as $\rm RM_{obs}$.
This may be due to differences in the polarization sources.
FD is the amount of rotation that polarized waves from a background source experience as they pass through the galaxy. 
In contrast, $\rm RM_{obs}$ corresponds to the rotation of polarized waves emitted from the galaxy itself.
Naively, we may guess that polarization originating from behind the position where the maximum polarized intensity is produced, i.e., the position where the magnetic fields are strongest, does not affect the amount obtained by integration along the LOS.
In figure \ref{fig:rm3}, we therefore show the FD integrated from the position of maximum polarization toward the observer.
This result shows that the overall absolute value of this integral is about half the value of the FD and is close to the value of $\rm RM_{obs}$.
Also, the spatial distribution is similar to that of $\rm RM_{obs}$.
The RM obtained from an actual observation can thus be considered to represent the FD from the position where the polarized radiation at the local position is strongest along the LOS.
Since it is difficult to determine from RM observations alone where a polarized radiation source is located along the LOS, one must consider methods to identify the three-dimensional distribution of the magnetic field.

\subsection{Depolarization}

\begin{figure}
    \begin{center}
        \includegraphics[width=0.8\linewidth]{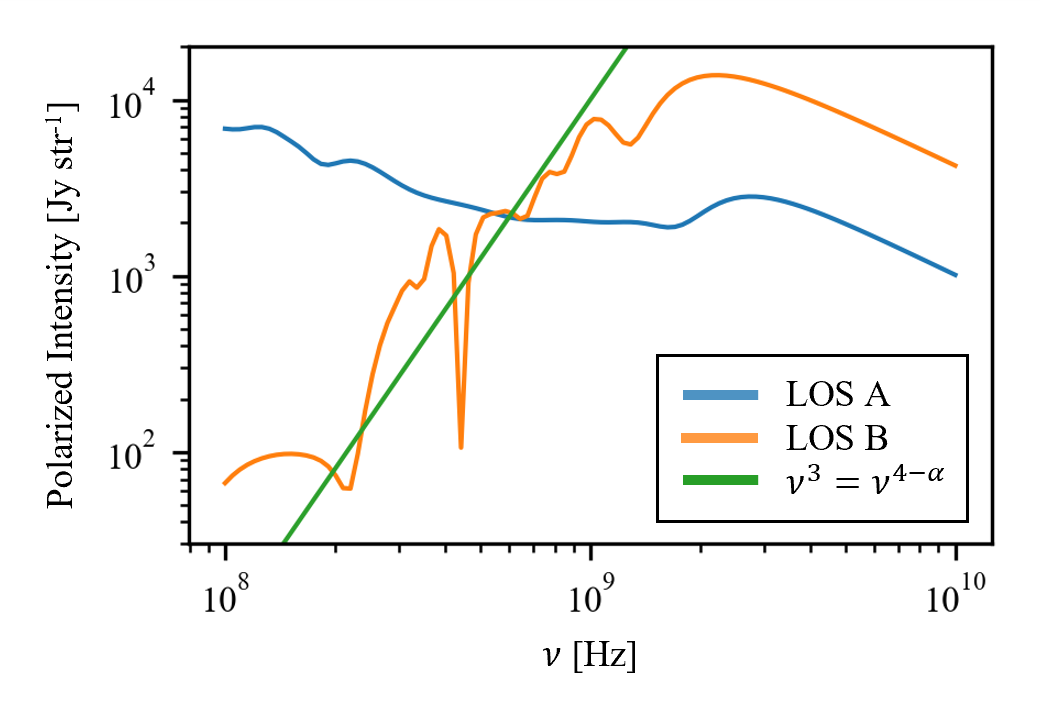}
    \end{center}
    \caption{
    The spectra of polarized intensity.
    The polarization intensities along LOS A (blue), which has a large depolarization effect at low frequencies, and along LOS B (orange), which has a small effect, are plotted.
    The green line shows the slope of the frequency dependence of internal depolarization B on the low frequency side, $p \propto \nu^{-3}$.
    LOS A: $(x,y)=(2 \mathrm{kpc}, -1 \mathrm{kpc})$, LOS B: $(x, y)=(4 \mathrm{kpc}, -5 \mathrm{kpc})$
    }
    \label{fig:sp_nu}
\end{figure}

\begin{figure}
    \begin{center}
    \includegraphics[width=0.8\linewidth]{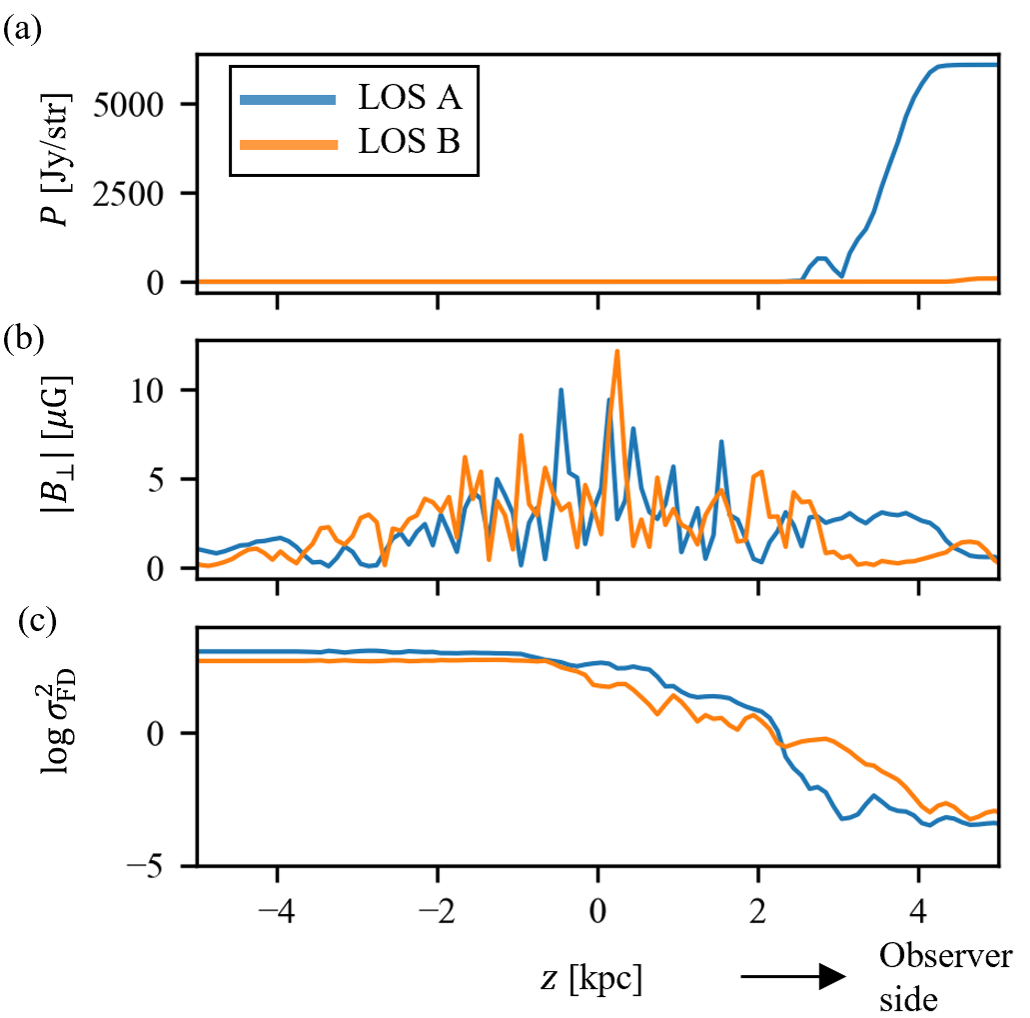}
    \end{center}
    \caption{
    (a) Changes in the integral of polarized intensity along LOS A (blue) and LOS B (orange).
    (b) Distribution of normalized magnetic field strengths perpendicular to the LOS on LOS A and LOS B.
    (c) Changes in the dispersion of the FD on LOS A and LOS B.
    The dispersion of the FD is larger far from the observer because of the greater length of the Faraday rotation integral.
    }
    \label{fig:b_z}
\end{figure}

Figure \ref{fig:sp_nu} shows the frequency dependence of the two lines of sight.
LOS A(2$\mathrm{\> kpc}$, $-1\mathrm{\> kpc}$) is the LOS that shows the highest polarized intensity in the pseudo-observation at 0.15$\mathrm{\> GHz}$.
In contrast, LOS B (4$\mathrm{\> kpc}$, $-5\mathrm{\> kpc}$) shows a strong polarized intensity at 8.5$\mathrm{\> GHz}$, but the polarized intensity is significantly reduced at 0.15$\mathrm{\> GHz}$.

At LOS A, the polarized intensity begins to decrease from around 2$\mathrm{\> GHz}$ due to the depolarization effect, but the polarized intensity increases again below about 0.5$\mathrm{\> GHz}$.
This is thought to occur because the position that produces the dominant radiation on the LOS has changed.
At LOS B, the polarized intensity begins to decrease below about 1$\mathrm{\> GHz}$ and continues to decrease significantly down to 0.15$\mathrm{\> GHz}$.
The green straight line is the frequency dependence of polarized intensity when the internal depolarization is strong.
The internal depolarization is known to follow a power law ($\propto\nu^{4-\alpha}$) when the Faraday dispersion is large \citep{arshakian_2011}.
Here, $\alpha$ is the synchrotron spectral index, which can be written as $\alpha=(s-1)/2$.
Because the slope of the spectrum of LOS B approximately matches the green line, this shows that the polarization is eliminated by internal depolarization.

Figure \ref{fig:b_z}a shows the relation between the length of the LOS and the polarized intensity at 0.15$\mathrm{\> GHz}$.
As in figure \ref{fig:sp_nu}, the blue and orange curves show the LOS positions where the polarized intensity is at its maximum and minimum, respectively, at this frequency.
The magnetic field strength perpendicular to each LOS is shown in figure \ref{fig:b_z}b.
LOS A and LOS B have similar magnetic field distributions within $|z| < 2\mathrm{\> kpc}$, but there is a difference for $z > 2.5\mathrm{\> kpc}$, where a strong magnetic field is seen only along LOS A.
Figure \ref{fig:b_z}c shows the FD dispersion.
Both LOS A and B have a strong perpendicular magnetic field at $z<2\mathrm{\> kpc}$, but the FD dispersion is also large, so the radiation is completely depolarized.
The dispersion along LOS A decreases suddenly for $z >2.5\mathrm{\> kpc}$, even though the perpendicular field remains high.
This implies that LOS A has an ordered magnetic field that is sufficiently large to cause polarized emission.
On the other hand, along LOS B, although the amplitude of the perpendicular magnetic field is reduced, turbulent fields have developed, and dispersion plays a role in depolarization. 
Therefore, the polarized emission along LOS A originates in the ordered magnetic field in the halo region.

\subsection{Comparison with actual observations}
\begin{figure}
    \begin{center}
    \includegraphics[width=0.9\linewidth]{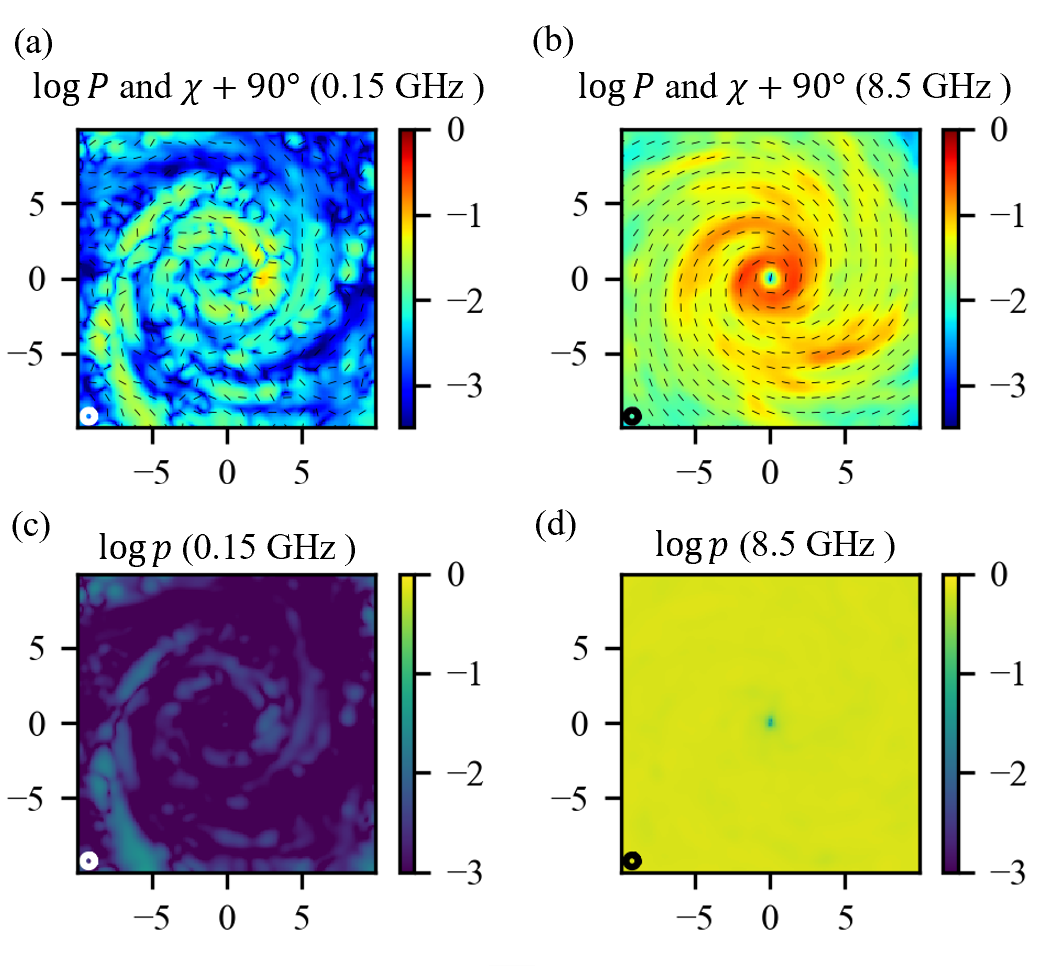}
    \end{center}
    \caption{
    Same as figure \ref{fig:sp_1_10}, but with the size of the observing beam changed.
    The Gaussian widths of the beams in these panels are 800 pc.
    The circle at the lower left in each panel shows the beam size.
    The polarized intensity has the units [$\mathrm{mJy\>beam^{-1}}$].
    }
    \label{fig:spc_1g}
\end{figure}

\begin{figure}
    \begin{center}
    \includegraphics[width=0.8\linewidth]{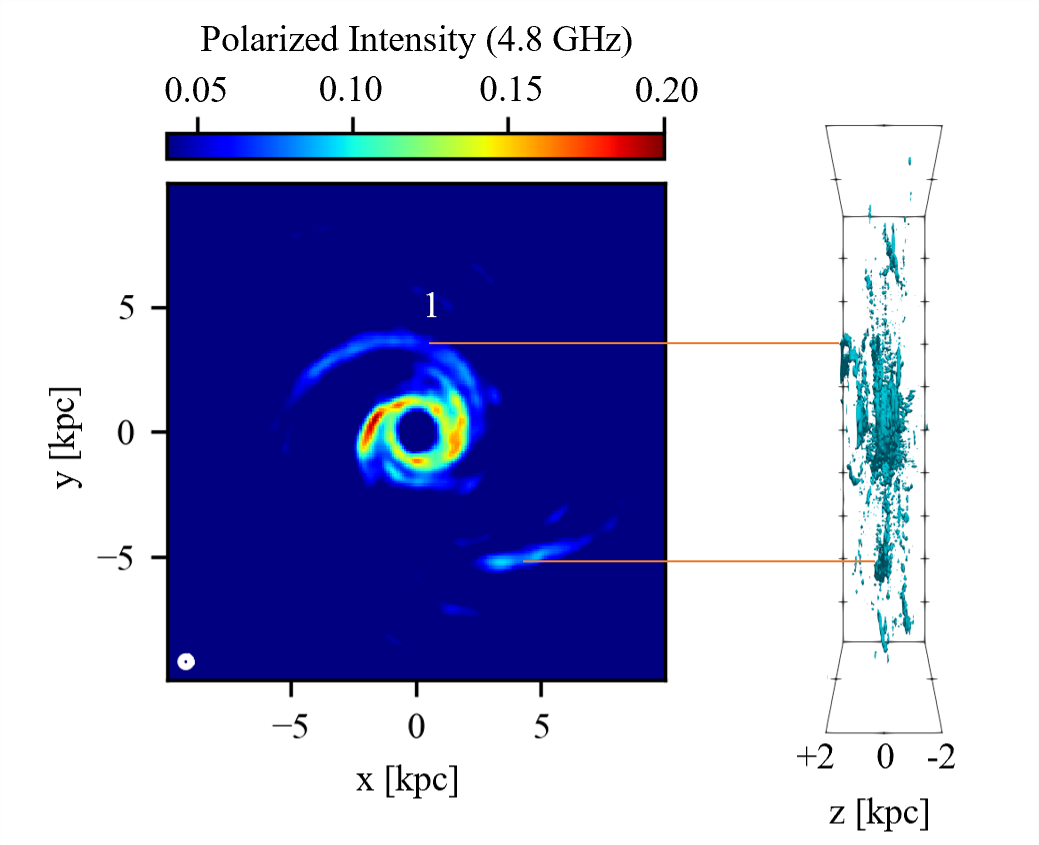}
    \end{center}
    \caption{
    (left) Polarized intensity map for a face-on spiral galaxy at the observing frequency 4.8$\mathrm{\> GHz}$.
    Assuming this to be an observation of IC\,342, the distance is 3.3$\mathrm{\> kpc}$, and the beam size is 25 arcsec.
    The units of the color bar are [$\rm mJy\>beam^{-1}$].
    (right) Three-dimensional distribution of the polarized radiation reaching the observer.
    The orange lines associate arm structures with the originating locations of the polarized radiation.
    }
    \label{fig:3d_5g}
\end{figure}

\begin{figure}
    \begin{center}
    \includegraphics[width=0.8\linewidth]{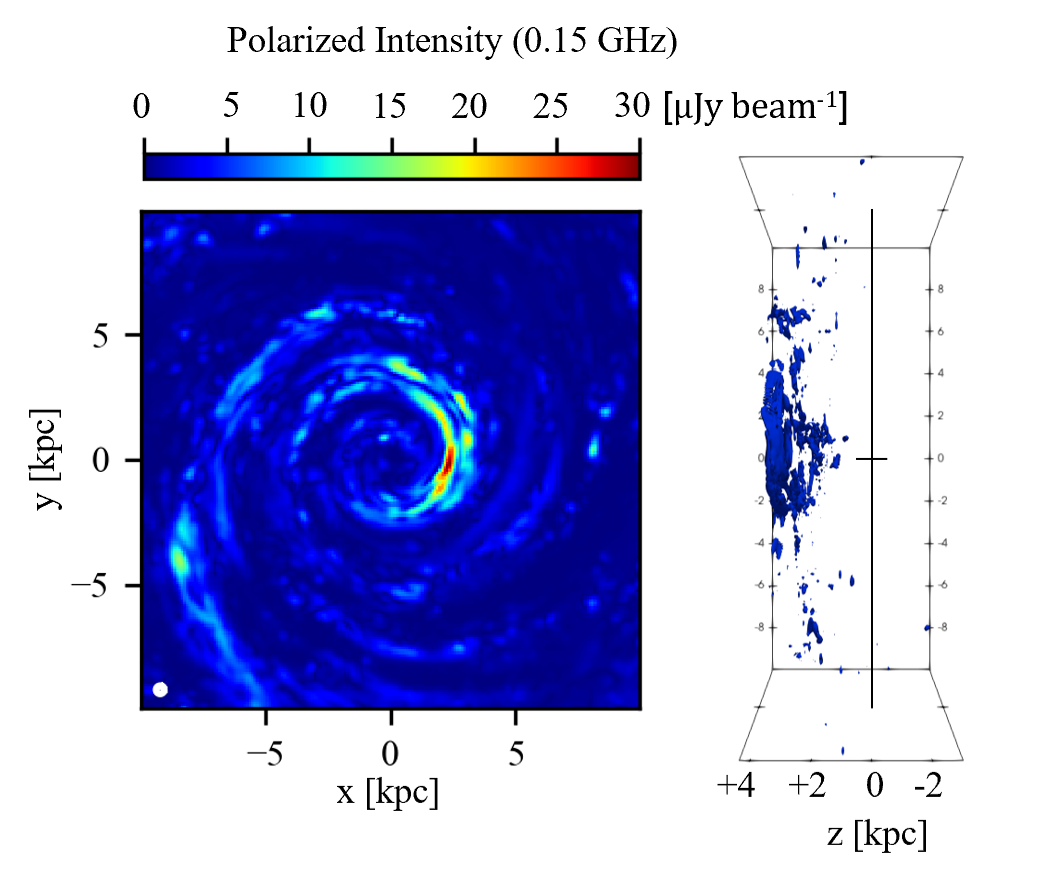}
    \end{center}
    \caption{
    The same as figure \ref{fig:3d_5g}, but at the observing frequency $0.15\mathrm{\>GHz}$. The units of the color bar are [$\rm \mu Jy\>beam^{-1}$].
    }
    \label{fig:3d_015g}
\end{figure}

In this section, we compare the results of our pseudo-observations with actual observations. Because of the superposition of polarized waves in the beam, the depolarization depends strongly on the beam size.
Figure \ref{fig:spc_1g} shows the polarized intensity and polarization degree at 0.15$\mathrm{\> GHz}$ and 8.5$\mathrm{\> GHz}$ with a Gaussian beamwidth of 800$\mathrm{\> pc}$.
The beam-averaged polarized intensity at 8.5$\mathrm{\> GHz}$ displayed in figure \ref{fig:spc_1g}d looks smoother than that in figure \ref{fig:sp_1_10}, but there is no significant change in either the polarization or the intensity distribution.
On the other hand, at 0.15 GHz, the degree of polarization $p$ is significantly reduced because the polarization angles are random.
As a result, $p<$ 0.1\% for most pixels.
To validate our pseudo-observations, we rescaled the beam size to that used for observations of the spiral galaxy IC\,342 (figure \ref{fig:3d_5g}).
We set the distance from the observer to be 3.3 Mpc and calculated the spatial resolution assuming $25"$ (400$\mathrm{\> pc}$).
The left panel of figure \ref{fig:3d_5g} shows a projection map of the polarized intensity at 4.8$\mathrm{\> GHz}$.
The polarization intensities obtained in our pseudo-observations reproduce well the features of the actual observations of IC\,342 \citep{beck_2015}.
The right panel of figure \ref{fig:3d_5g} shows the three-dimensional distribution of the polarized intensity, which enables us to understand where the projected polarized radiation originates in the galactic disk.
The green surfaces are isosurfaces of the polarized intensity at $0.08 \mathrm{\>Jy\>beam^{-1}}$; they show that the emission region extends over $\pm 2\mathrm{\> kpc}$ from the equatorial plane.
The orange lines show the correspondence between the obtained arm structure and the radiation position, which show that each arm is located at a different height.
For example, we can see that two or more spiral fields overlap in arm 1 in figure \ref{fig:3d_5g}. Even though the spiral arms in the projection appear to be a single structure, they are not located on the same slice in the height direction.
In the pseudo-observations, a region with strong polarization exists at $|x|<2.5\mathrm{\> kpc}$, $|y|<2.5\mathrm{\> kpc}$, which can be seen even at 1$\mathrm{\> GHz}$. 
However, in the actual observations, this feature disappears due to depolarization at about 1.4$\mathrm{\> GHz}$ \citep{beck_2015}.
This occurs because the model of \cite{machida_2013} neglects the cold HI clouds that produce turbulent magnetic fields with short length scales in the disk.
If we were to treat this cold part of the disk, the depolarization due to Faraday rotation would be more effective, and the polarized emission would be reduced \citep{borlaff_2021}.

Next, we consider depolarization effects using pseudo-observations at 0.15 GHz (figure \ref{fig:3d_015g}).
We assumed that these are observations of IC 342 at a resolution of 25 arcseconds.
The maximum polarization intensity at 0.15 GHz is weaker than at 4.8 GHz, reaching only about 30$\mathrm{\>\mu Jy\>beam^{-1}}$.
This is because the polarized intensity produced by the disk is depolarized, and only the dilute halo component remains in the integrated intensity.
In addition, the correlation length of magnetic turbulence in the present model galaxy tends to be long because only the high-temperature plasma in the galactic disk is included.
The current model assumes that we are dealing with warm and hot plasma.
However, real galactic gas disks form high-density, cold gas disks, and these regions have strong magnetic fields.
In fact, the central region in Figure \ref{fig:3d_015g} where the polarization intensity shows a maximum value should be composed of cold gas. 
Such a high-density region would be expected to have stronger depolarization than the current model.
Therefore, in order to calculate the depolarization effect correctly, a multi-temperature disk must be taken into account.
Compared with the three-dimensional structure of polarized radiation shown in Figure \ref{fig:3d_5g}, the right panel of figure \ref{fig:3d_015g} shows that the magnetic field structure of the halo is being observed, as shown in the previous section.
If LOFAR were used to observe with the NL-remote, even 3$\sigma$ detection of such weak halo magnetic fields requires an observation time of $10^5$ hours or more \citep{haarlem_2013}.
SKA1-Low has a resolution of about 5 arcsec at 150 MHz.
Halo radiation at this resolution is about 2 $\mu$Jy/beam.
To detect 3$\sigma$ of this radiation with 100 MHz bandwidth observation requires about 80 hours of observation time.
(SKA Baseline Design document version2)
\footnote{SKA Baseline Design document version 2, 2015
$\langle$https://www.skatelescope.org/key-documents/$\rangle$.}.
The sensitivity of SKA2 is about 10 times better than that of SKA1.
Therefore, using SKA2, even with 10 MHz bandwidth observation, 5$\sigma$ detection is possible in a few tens hours of observation time.

\begin{figure}
    \begin{center}
        \includegraphics[width=0.8\linewidth]{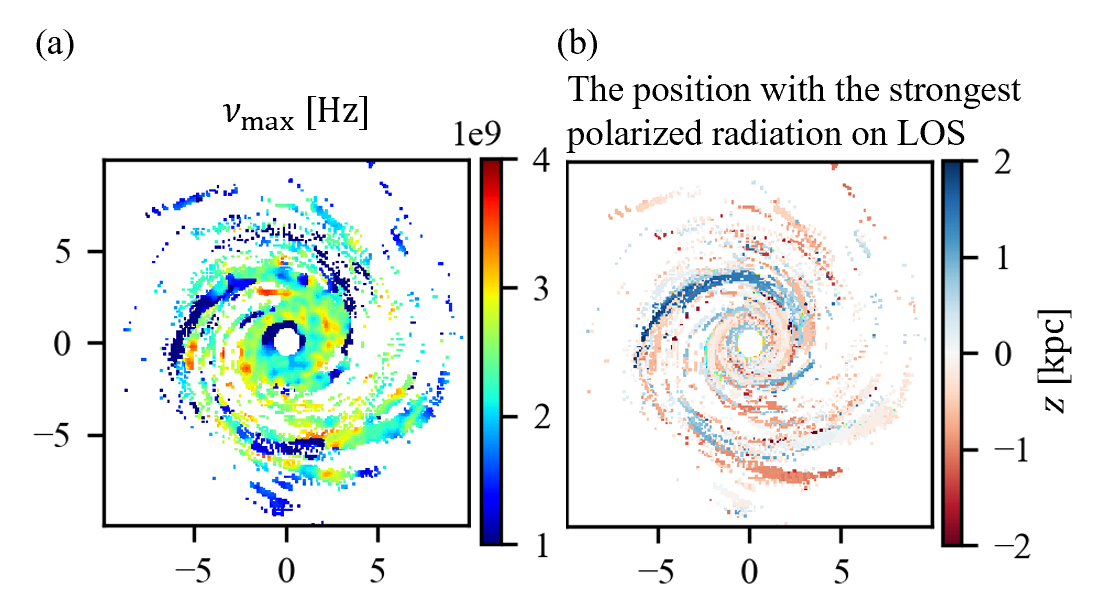}
    \end{center}
    \caption{
        (a) Map of the observing frequency that maximizes the intensity of polarized radiation.
        (b) Map showing the positions from which the strongest polarized radiation is emitted along the LOS at 8.5 GHz.
        }
    \label{fig:nu_map}
\end{figure}

\begin{figure}
    \begin{center}
        \includegraphics[width=0.8\linewidth]{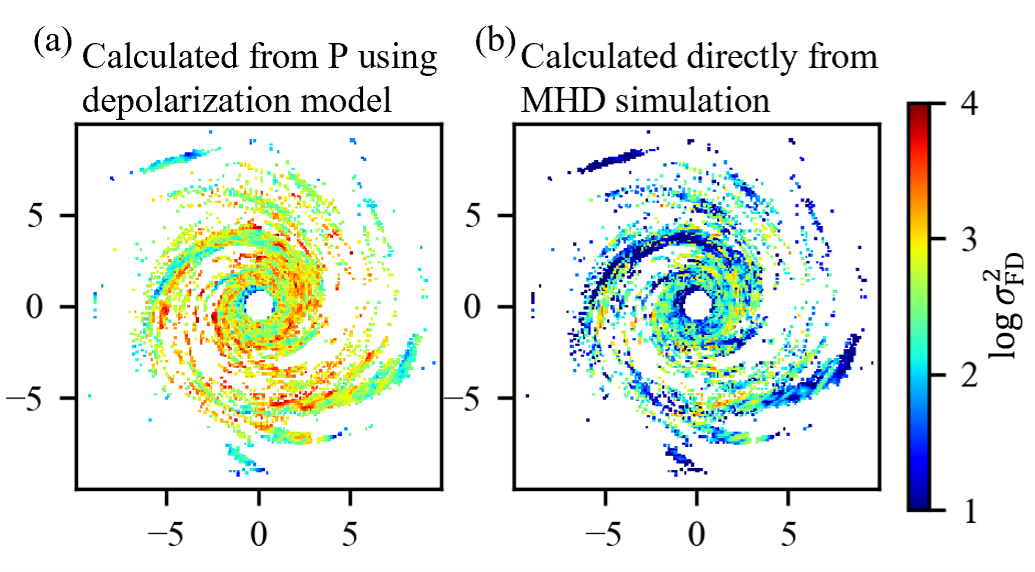}
    \end{center}
    \caption{
    (a) Map of FD dispersion obtained from the frequency dependence of the polarized intensity using the internal depolarization model.
    (b) Map of FD dispersion between the observer and the position where the polarized radiation is the strongest, which is derived directly from the MHD simulation.
    }
    \label{fig:sigma_2}
\end{figure}

We next consider whether it is possible to retrieve information about the spatial distribution along the LOS from the polarized intensity.
The polarized intensity shows a maximum at the frequencies where beam depolarization begins to operate (e.g., figure \ref{fig:sp_nu}, \cite{arshakian_2011}).
This feature has the potential to be detectable at locations along the LOS.
A region with a peak at a low frequency means that there is no turbulent and dense plasma in front of it, and it is expected to be distributed in front of the galactic disk, where it is more susceptible to depolarization.
In figure \ref{fig:nu_map}a, we therefore plot the frequency at which the polarized intensity becomes the strongest along the LOS, and we show the corresponding position in figure \ref{fig:nu_map}b.
The color bar in figure \ref{fig:nu_map}b represents the $z$ coordinate, with positive being the observer's side.
The spiral structure extends over the range $y = 0-5$ kpc, with an emitting region at $z = 2$ kpc on the observer's side.
A similar spiral can be seen near $y=-2$ kpc and $x=0-4$ kpc.
However, although the arm structure near $y = -5$ kpc has a part with a low frequency peak, the actual radiating position is near the equatorial plane.
The difference between these correlations can be determined from the length scale of the peak frequency map.
When the frequency peak is mottled, it is influenced by the random magnetic field in the disk.
On the other hand, a part where the length scale of the peak frequency structure is long reflects the structure of the magnetic flux tube above the disk.

Since the peak frequency depends on the dispersion of the FD, we obtained that quantity using the dispersion model and utilized the numerical results to check the verification.
Figures \ref{fig:sigma_2}a and \ref{fig:sigma_2}b are FD dispersion maps computed from the internal depolarization model (see Equation \ref{eq:DA}) and from the numerical simulation (Equation \ref{eq:sigma_FD}), respectively.
The FD dispersion of the front spiral arm ($y=1$--5) shown in figure \ref{fig:sigma_2}b shows that the values are more than one order of magnitude smaller than the values of $\sigma^2_\mathrm{FD}$ around the spiral arm for both derivation methods.
Although the density of a galactic halo is 10--100 times lower than the density of a galactic disk, magnetic flux tubes with lengths greater than the turbulence length scale may be the origin of the polarized emission.
Comparing the absolute values of the FD dispersion, we find that the results obtained using the depolarization model (figure \ref{fig:sigma_2}a) are higher than those obtained from the direct calculation (figure \ref{fig:sigma_2}b). This means that the turbulent magnetic field may be overestimated from the observations.

\subsection{Validity of the pseudo-observation assumptions}

\begin{figure}
    \begin{center}
    \includegraphics[width=0.8\linewidth]{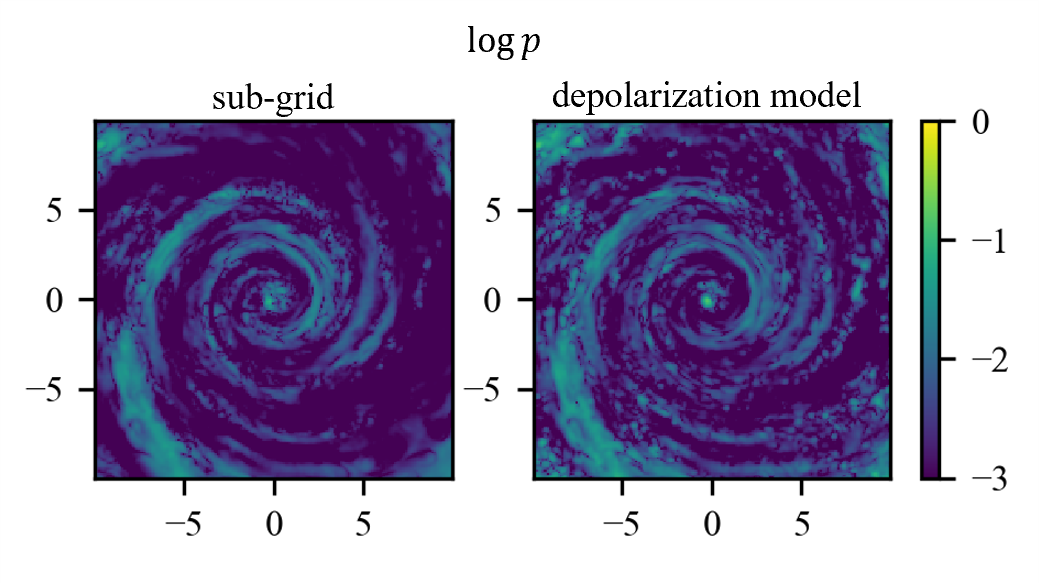}
    \end{center}
    \caption{
    Maps of the degree of polarization at the observing frequency 0.15$\mathrm{\> GHz}$. Both maps have the same beam size.
    (left) Calculated by applying sub-grid turbulence.
    (right) Calculated by our method using depolarization models.
    }
    \label{fig:kensho}
\end{figure}

The total intensities obtained from pseudo-observations using cosmological galaxy simulations have been reported to be lower than the actual observations when cosmic-ray electron energies are assumed to be equivalent to the magnetic energy \citep{reissl_2019}. 
Since our simulation also assumes equipartition of the energy, the trend of the pseudo-observation is similar to the previous results.
On the other hand, the polarized intensity in the simulation agrees with the actual intensity.
This suggests that, although the mean field that originates the polarized emission is reproduced, the current numerical simulations do not provide enough turbulent field to explain the total intensity.
Observational evidence shows that the energy in cosmic-ray electrons is about 50 times smaller than that in cosmic-ray ions \citep{2016ApJ...831...18C}.
Since the energy in cosmic-ray electrons is balanced by the mean magnetic field, the energy in invisible cosmic-ray ions may be higher.
Also, the total energy of the magnetic field may be underestimated by the amount of invisible cosmic-ray ions.
Since cosmic-ray ions enhance magnetic turbulence \citep{kuwabara_2004, kuwabara_2015}, a magnetic instability may easily be able to amplify the turbulent field to the required energy.
It will be necessary to simulate the dynamics of cosmic rays with MHD flows to obtain more accurate observables.
The galactic disk contains a mixture of multi-temperature structures such as molecular clouds, HI regions, and high-temperature plasma, each of which generates turbulence with its own length scale \citep{inutsuka_2015}.
In each component, the turbulent magnetic energy is comparable to the gas pressure, and low-temperature components are embedded in the high-temperature components.
These various components are also important as a cause of magnetic turbulence below the resolution limit.

To verify this method of using depolarization models, we compared our pseudo-observation code with the method used by \cite{reissl_2019} to apply a turbulent sub-grid magnetic field and perform beam averaging.
Figure \ref{fig:kensho} shows the polarization degree at 146 MHz obtained using the two methods.
The left panel of figure \ref{fig:kensho} shows the result obtained by applying sub-grid turbulence, and the right panel shows the result obtained in this study using depolarization models.
The sub-grid width is 1/3 of the original grid width in each direction, so the numerical cost of the sub-grid method is 9 times larger than that of our method.
There was almost no difference in the degree of polarization, which shows that the calculations using depolarization models give reasonable results.
We have thus shown that Burn's depolarization model successfully reproduces internal depolarization when the turbulence scale is smaller than the MHD grid size.

The right panel of figure \ref{fig:kensho} corresponds to the case where the length scale of the turbulent magnetic field in figure \ref{fig:sp_1_10}c is reduced by the factor 1/3.
Although the turbulent length scale is shorter than in figure \ref{fig:sp_1_10}c, the spatial distribution shows no significant difference.
In other words, a reasonable value for the length scale of the turbulent magnetic field is about 100 pc. 

\section{Summary}
We have performed pseudo-observations using physical quantities obtained from MHD simulations in order to clarify the relationship between the projected intensity and the emitting region in real space.
We have also focused on the importance of depolarization at meter wavelengths.
Here, we summarize our findings:

\begin{itemize}
    \item Assuming a linear relationship between polarization angle and wavelength squared, we found that the value of the RM is the amount of Faraday rotation obtained from the brightest polarized emission region along the LOS to the observer.
    This value may underestimate the strength of the magnetic field.

    \item Depolarization depends strongly on the properties of the turbulence, so it is necessary to handle the turbulent magnetic field carefully. When wavelength-independent polarization is calculated assuming isotropic turbulence, the effect of depolarization is overestimated because most of the polarized intensity is emitted from the surroundings of the spiral arm, and the turbulent magnetic field along the magnetic spiral arms has strong inhomogeneity \citep{machida_2019}.

    \item Internal depolarization can be divided into two parts: depolarization that occurs within the emitting region and depolarization that occurs from the emitting region to the observer.
    We found that the latter produces the most efficient depolarization of all the depolarization models.

    \item Polarized emission in the 0.15$\mathrm{\> GHz}$ band originates from the magnetic field of the halo, since emission from the equatorial plane is reduced by strong depolarization.
    This halo emission is difficult to detect with current facilities. 
    But it may be detectable if sufficient observation time is obtained using SKA1.
    Also, SKA2 is expected to be sufficient to detect these structures.

    \item Since the projected intensity originates from several emitting regions located at different heights in real space, it is necessary to pay attention to the correlation between the spatial distribution of the emitting filaments and the projected intensity map.
    By investigating the frequency at which beam depolarization begins to operate, we found that it is possible to separate emitting regions with different heights.
    
\end{itemize}

The present model does not take into account cooling, which is important for the formation of molecular clouds and HI clouds.
When cooling is neglected, the coherence length of the turbulence increases because the thermal energy is greater.
This means that depolarization will be weaker than is the actual case for both wavelength-independent and wavelength-dependent depolarization.
Recent high-resolution, far-infrared polarimetric observations of M\,51 have identified a magnetic field with a coherence length along the spiral arms that is shorter than that of the magnetic field derived from radio observations \citep{borlaff_2021}.
Several groups, including ours, have already stared to conduct 3D MHD simulations for the galactic gas disk, including radiative cooling \citep{reissl_2019,kudoh_2020}.
In the future, we will consider pseudo-observations that take into account results for cold galactic disks.

\bigskip
\begin{ack}
We are grateful to Drs. H. Sakemi, M. Haverkorn, S. Totorica, and Mr. R. Omae for useful discussion and  brushing up on our writing.
We also thank the anonymous referee for his/her useful comments and constructive suggestions.
This work was supported in part by JSPS KAKENHI (MM:19K03916, 20H01941, TO:22K14032).
Our numerical computations were carried out on SX-9 and on analysis servers at the Center for Computational Astrophysics of the National Astronomical Observatory of Japan.
The computation was carried out using the computer resources offered by the Research Institute for Information Technology, Kyushu University.
\end{ack}

\bibliographystyle{aasjournal.bst}
\bibliography{main}

\begin{thebibliography}{}
\expandafter\ifx\csname natexlab\endcsname\relax\def\natexlab#1{#1}\fi
\providecommand{\url}[1]{\href{#1}{#1}}

\bibitem[{{Akahori} {et~al.}(2018){Akahori}, {Nakanishi}, {Sofue}, {Fujita},
  {Ichiki}, {Ideguchi}, {Kameya}, {Kudoh}, {Kudoh}, {Machida}, {Miyashita},
  {Ohno}, {Ozawa}, {Takahashi}, {Takizawa}, \& {Yamazaki}}]{akahori_2018}
{Akahori}, T., {Nakanishi}, H., {Sofue}, Y., {et~al.} 2018, \pasj, 70, R2

\bibitem[{{Arshakian} \& {Beck}(2011)}]{arshakian_2011}
{Arshakian}, T.~G., \& {Beck}, R. 2011, \mnras, 418, 2336

\bibitem[{{Beck}(2015{\natexlab{a}})}]{beck_2015}
{Beck}, R. 2015{\natexlab{a}}, \aap, 578, A93

\bibitem[{{Beck}(2015{\natexlab{b}})}]{beck_2015_m}
---. 2015{\natexlab{b}}, \aapr, 24, 4

\bibitem[{{Birnboim} {et~al.}(2015){Birnboim}, {Balberg}, \&
  {Teyssier}}]{birnboim_2015}
{Birnboim}, Y., {Balberg}, S., \& {Teyssier}, R. 2015, \mnras, 447, 3678

\bibitem[{{Borlaff} {et~al.}(2021){Borlaff}, {Lopez-Rodriguez}, {Beck},
  {Stepanov}, {Ntormousi}, {Hughes}, {Tassis}, {Marcum}, {Grosset}, {Beckman},
  {Proudfit}, {Clark}, {D{\'\i}az-Santos}, {Mao}, {Reach}, {Roman-Duval},
  {Subramanian}, {Tram}, {Zweibel}, {Dale}, \& {Legacy Team}}]{borlaff_2021}
{Borlaff}, A.~S., {Lopez-Rodriguez}, E., {Beck}, R., {et~al.} 2021, \apj, 921,
  128

\bibitem[{{Brentjens} \& {de Bruyn}(2005)}]{brentjens_2005}
{Brentjens}, M.~A., \& {de Bruyn}, A.~G. 2005, \aap, 441, 1217

\bibitem[{{Burn}(1966)}]{burn_1966}
{Burn}, B.~J. 1966, \mnras, 133, 67

\bibitem[{{Cummings} {et~al.}(2016){Cummings}, {Stone}, {Heikkila}, {Lal},
  {Webber}, {J{\'o}hannesson}, {Moskalenko}, {Orlando}, \&
  {Porter}}]{2016ApJ...831...18C}
{Cummings}, A.~C., {Stone}, E.~C., {Heikkila}, B.~C., {et~al.} 2016, \apj, 831,
  18

\bibitem[{{Fletcher}(2010)}]{fletcher_2010}
{Fletcher}, A. 2010, in Astronomical Society of the Pacific Conference Series,
  Vol. 438, The Dynamic Interstellar Medium: A Celebration of the Canadian
  Galactic Plane Survey, ed. R.~{Kothes}, T.~L. {Landecker}, \& A.~G. {Willis},
  197

\bibitem[{{Fletcher} {et~al.}(2011){Fletcher}, {Beck}, {Shukurov},
  {Berkhuijsen}, \& {Horellou}}]{fletcher_2011}
{Fletcher}, A., {Beck}, R., {Shukurov}, A., {Berkhuijsen}, E.~M., \&
  {Horellou}, C. 2011, \mnras, 412, 2396

\bibitem[{{Hanasz} {et~al.}(2009){Hanasz}, {W{\'o}lta{\'n}ski}, \&
  {Kowalik}}]{hanasz_2009}
{Hanasz}, M., {W{\'o}lta{\'n}ski}, D., \& {Kowalik}, K. 2009, \apjl, 706, L155

\bibitem[{{Inutsuka} {et~al.}(2015){Inutsuka}, {Inoue}, {Iwasaki}, {Stone},
  {Suzuki}, {Tsukamoto}, \& {Takamoto}}]{inutsuka_2015}
{Inutsuka}, S.~i., {Inoue}, T., {Iwasaki}, K., {et~al.} 2015, in Astronomical
  Society of the Pacific Conference Series, Vol. 498, Numerical Modeling of
  Space Plasma Flows ASTRONUM-2014, ed. N.~V. {Pogorelov}, E.~{Audit}, \& G.~P.
  {Zank}, 75

\bibitem[{{Kierdorf} {et~al.}(2020){Kierdorf}, {Mao}, {Beck}, {Basu},
  {Fletcher}, {Horellou}, {Tabatabaei}, {Ott}, \& {Haverkorn}}]{kierdorf_2020}
{Kierdorf}, M., {Mao}, S.~A., {Beck}, R., {et~al.} 2020, \aap, 642, A118

\bibitem[{{Krause}(2009)}]{krause_2009}
{Krause}, M. 2009, in Revista Mexicana de Astronomia y Astrofisica Conference
  Series, Vol.~36, Revista Mexicana de Astronomia y Astrofisica Conference
  Series, 25--29

\bibitem[{{Kudoh} {et~al.}(2020){Kudoh}, {Wada}, \& {Norman}}]{kudoh_2020}
{Kudoh}, Y., {Wada}, K., \& {Norman}, C. 2020, \apj, 904, 9

\bibitem[{{Kurahara} {et~al.}(2021){Kurahara}, {Nakanishi}, \&
  {Kudoh}}]{kurahara_2021}
{Kurahara}, K., {Nakanishi}, H., \& {Kudoh}, Y. 2021, \pasj, 73, 220

\bibitem[{{Kuwabara} \& {Ko}(2015)}]{kuwabara_2015}
{Kuwabara}, T., \& {Ko}, C.-M. 2015, \apj, 798, 79

\bibitem[{{Kuwabara} {et~al.}(2004){Kuwabara}, {Nakamura}, \&
  {Ko}}]{kuwabara_2004}
{Kuwabara}, T., {Nakamura}, K., \& {Ko}, C.~M. 2004, \apj, 607, 828

\bibitem[{{Machida} {et~al.}(2018){Machida}, {Akahori}, {Nakamura},
  {Nakanishi}, \& {Haverkorn}}]{machida_2018}
{Machida}, M., {Akahori}, T., {Nakamura}, K.~E., {Nakanishi}, H., \&
  {Haverkorn}, M. 2018, \mnras, 480, 17

\bibitem[{{Machida} {et~al.}(2019){Machida}, {Akahori}, {Nakamura},
  {Nakanishi}, \& {Haverkorn}}]{machida_2019}
---. 2019, \mnras, 482, 3394

\bibitem[{{Machida} {et~al.}(2013){Machida}, {Nakamura}, {Kudoh}, {Akahori},
  {Sofue}, \& {Matsumoto}}]{machida_2013}
{Machida}, M., {Nakamura}, K.~E., {Kudoh}, T., {et~al.} 2013, \apj, 764, 81

\bibitem[{{Machida} {et~al.}(2009){Machida}, {Matsumoto}, {Nozawak},
  {Takahashi}, {Fukui}, {Kudo}, {Torii}, {Yamamoto}, {Fujishita}, \&
  {Tomisaki}}]{machida_2009}
{Machida}, M., {Matsumoto}, R., {Nozawak}, S., {et~al.} 2009, \pasj, 61, 411

\bibitem[{{Mao} {et~al.}(2012){Mao}, {McClure-Griffiths}, {Gaensler},
  {Haverkorn}, {Beck}, {McConnell}, {Wolleben}, {Stanimirovi{\'c}}, {Dickey},
  \& {Staveley-Smith}}]{mao_2012}
{Mao}, S.~A., {McClure-Griffiths}, N.~M., {Gaensler}, B.~M., {et~al.} 2012,
  \apj, 759, 25

\bibitem[{{Miyamoto} \& {Nagai}(1975)}]{miyamoto_1975}
{Miyamoto}, M., \& {Nagai}, R. 1975, \pasj, 27, 533

\bibitem[{{Mulcahy} {et~al.}(2014){Mulcahy}, {Horneffer}, {Beck}, {Heald},
  {Fletcher}, {Scaife}, {Adebahr}, {Anderson}, {Bonafede}, {Br{\"u}ggen},
  {Brunetti}, {Chy{\.z}y}, {Conway}, {Dettmar}, {En{\ss}lin}, {Haverkorn},
  {Horellou}, {Iacobelli}, {Israel}, {Junklewitz}, {Jurusik}, {K{\"o}hler},
  {Kuniyoshi}, {Orr{\'u}}, {Paladino}, {Pizzo}, {Reich}, \&
  {R{\"o}ttgering}}]{mulcahy_2014}
{Mulcahy}, D.~D., {Horneffer}, A., {Beck}, R., {et~al.} 2014, \aap, 568, A74

\bibitem[{{Nishikori} {et~al.}(2006){Nishikori}, {Machida}, \&
  {Matsumoto}}]{nishikori_2006}
{Nishikori}, H., {Machida}, M., \& {Matsumoto}, R. 2006, \apj, 641, 862

\bibitem[{{Ohno} \& {Shibata}(1993)}]{ohno_1993}
{Ohno}, H., \& {Shibata}, S. 1993, \mnras, 262, 953

\bibitem[{{O'Sullivan} \& {Gabuzda}(2009)}]{o'sullivan_2009}
{O'Sullivan}, S.~P., \& {Gabuzda}, D.~C. 2009, \mnras, 393, 429

\bibitem[{{O'Sullivan} {et~al.}(2012){O'Sullivan}, {Brown}, {Robishaw},
  {Schnitzeler}, {McClure-Griffiths}, {Feain}, {Taylor}, {Gaensler},
  {Landecker}, {Harvey-Smith}, \& {Carretti}}]{o'sullivan_2012}
{O'Sullivan}, S.~P., {Brown}, S., {Robishaw}, T., {et~al.} 2012, \mnras, 421,
  3300

\bibitem[{{Pacholczyk}(1970)}]{pacholczyk_1970}
{Pacholczyk}, A.~G. 1970, Radio astrophysics : nonthermal processes in galactic
  and extragalactic sources (San Francisco (Calif.) : Freeman)

\bibitem[{{Pakmor} {et~al.}(2014){Pakmor}, {Marinacci}, \&
  {Springel}}]{pakmor_2014}
{Pakmor}, R., {Marinacci}, F., \& {Springel}, V. 2014, \apjl, 783, L20

\bibitem[{{Rand} \& {Kulkarni}(1989)}]{rand_1989}
{Rand}, R.~J., \& {Kulkarni}, S.~R. 1989, \apj, 343, 760

\bibitem[{{Reissl} {et~al.}(2019){Reissl}, {Brauer}, {Klessen}, \&
  {Pellegrini}}]{reissl_2019}
{Reissl}, S., {Brauer}, R., {Klessen}, R.~S., \& {Pellegrini}, E.~W. 2019,
  \apj, 885, 15

\bibitem[{{Sarazin}(1999)}]{sarazin_1999}
{Sarazin}, C.~L. 1999, \apj, 520, 529

\bibitem[{{Shneider} {et~al.}(2014{\natexlab{a}}){Shneider}, {Haverkorn},
  {Fletcher}, \& {Shukurov}}]{shneider_2014a}
{Shneider}, C., {Haverkorn}, M., {Fletcher}, A., \& {Shukurov}, A.
  2014{\natexlab{a}}, \aap, 567, A82

\bibitem[{{Shneider} {et~al.}(2014{\natexlab{b}}){Shneider}, {Haverkorn},
  {Fletcher}, \& {Shukurov}}]{shneider_2014b}
---. 2014{\natexlab{b}}, \aap, 568, A83

\bibitem[{{Soida} {et~al.}(2011){Soida}, {Krause}, {Dettmar}, \&
  {Urbanik}}]{soida_2011}
{Soida}, M., {Krause}, M., {Dettmar}, R.~J., \& {Urbanik}, M. 2011, \aap, 531,
  A127

\bibitem[{{Sokoloff} {et~al.}(1998){Sokoloff}, {Bykov}, {Shukurov},
  {Berkhuijsen}, {Beck}, \& {Poezd}}]{sokoloff_1998}
{Sokoloff}, D.~D., {Bykov}, A.~A., {Shukurov}, A., {et~al.} 1998, \mnras, 299,
  189

\bibitem[{{Springel} {et~al.}(2008){Springel}, {Wang}, {Vogelsberger},
  {Ludlow}, {Jenkins}, {Helmi}, {Navarro}, {Frenk}, \& {White}}]{springel_2008}
{Springel}, V., {Wang}, J., {Vogelsberger}, M., {et~al.} 2008, \mnras, 391,
  1685

\bibitem[{{Stepanov} {et~al.}(2008){Stepanov}, {Arshakian}, {Beck}, {Frick}, \&
  {Krause}}]{stepanoov_2008}
{Stepanov}, R., {Arshakian}, T.~G., {Beck}, R., {Frick}, P., \& {Krause}, M.
  2008, \aap, 480, 45

\bibitem[{{Sun} {et~al.}(2008){Sun}, {Reich}, {Waelkens}, \&
  {En{\ss}lin}}]{sun_2008}
{Sun}, X.~H., {Reich}, W., {Waelkens}, A., \& {En{\ss}lin}, T.~A. 2008, \aap,
  477, 573

\bibitem[{{van Haarlem} {et~al.}(2013){van Haarlem}, {Wise}, {Gunst}, {Heald},
  {McKean}, {Hessels}, {de Bruyn}, {Nijboer}, {Swinbank}, {Fallows},
  {Brentjens}, {Nelles}, {Beck}, {Falcke}, {Fender}, {H{\"o}randel},
  {Koopmans}, {Mann}, {Miley}, {R{\"o}ttgering}, {Stappers}, {Wijers},
  {Zaroubi}, {van den Akker}, {Alexov}, {Anderson}, {Anderson}, {van Ardenne},
  {Arts}, {Asgekar}, {Avruch}, {Batejat}, {B{\"a}hren}, {Bell}, {Bell}, {van
  Bemmel}, {Bennema}, {Bentum}, {Bernardi}, {Best}, {B{\^\i}rzan}, {Bonafede},
  {Boonstra}, {Braun}, {Bregman}, {Breitling}, {van de Brink}, {Broderick},
  {Broekema}, {Brouw}, {Br{\"u}ggen}, {Butcher}, {van Cappellen}, {Ciardi},
  {Coenen}, {Conway}, {Coolen}, {Corstanje}, {Damstra}, {Davies}, {Deller},
  {Dettmar}, {van Diepen}, {Dijkstra}, {Donker}, {Doorduin}, {Dromer}, {Drost},
  {van Duin}, {Eisl{\"o}ffel}, {van Enst}, {Ferrari}, {Frieswijk}, {Gankema},
  {Garrett}, {de Gasperin}, {Gerbers}, {de Geus}, {Grie{\ss}meier}, {Grit},
  {Gruppen}, {Hamaker}, {Hassall}, {Hoeft}, {Holties}, {Horneffer}, {van der
  Horst}, {van Houwelingen}, {Huijgen}, {Iacobelli}, {Intema}, {Jackson},
  {Jelic}, {de Jong}, {Juette}, {Kant}, {Karastergiou}, {Koers}, {Kollen},
  {Kondratiev}, {Kooistra}, {Koopman}, {Koster}, {Kuniyoshi}, {Kramer},
  {Kuper}, {Lambropoulos}, {Law}, {van Leeuwen}, {Lemaitre}, {Loose}, {Maat},
  {Macario}, {Markoff}, {Masters}, {McFadden}, {McKay-Bukowski}, {Meijering},
  {Meulman}, {Mevius}, {Middelberg}, {Millenaar}, {Miller-Jones}, {Mohan},
  {Mol}, {Morawietz}, {Morganti}, {Mulcahy}, {Mulder}, {Munk}, {Nieuwenhuis},
  {van Nieuwpoort}, {Noordam}, {Norden}, {Noutsos}, {Offringa}, {Olofsson},
  {Omar}, {Orr{\'u}}, {Overeem}, {Paas}, {Pandey-Pommier}, {Pandey}, {Pizzo},
  {Polatidis}, {Rafferty}, {Rawlings}, {Reich}, {de Reijer}, {Reitsma},
  {Renting}, {Riemers}, {Rol}, {Romein}, {Roosjen}, {Ruiter}, {Scaife}, {van
  der Schaaf}, {Scheers}, {Schellart}, {Schoenmakers}, {Schoonderbeek},
  {Serylak}, {Shulevski}, {Sluman}, {Smirnov}, {Sobey}, {Spreeuw}, {Steinmetz},
  {Sterks}, {Stiepel}, {Stuurwold}, {Tagger}, {Tang}, {Tasse}, {Thomas},
  {Thoudam}, {Toribio}, {van der Tol}, {Usov}, {van Veelen}, {van der Veen},
  {ter Veen}, {Verbiest}, {Vermeulen}, {Vermaas}, {Vocks}, {Vogt}, {de Vos},
  {van der Wal}, {van Weeren}, {Weggemans}, {Weltevrede}, {White}, {Wijnholds},
  {Wilhelmsson}, {Wucknitz}, {Yatawatta}, {Zarka}, {Zensus}, \& {van
  Zwieten}}]{haarlem_2013}
{van Haarlem}, M.~P., {Wise}, M.~W., {Gunst}, A.~W., {et~al.} 2013, \aap, 556,
  A2

\end{thebibliography}

\end{document}